\pdfoutput=1
\documentclass[aps,floatfix,10pt,twocolumn,prl]{revtex4-2}
\makeatletter
\def\NAT@sort{0}
\makeatother
\usepackage{mathtools,amsthm}
\newtheorem{theorem}{Theorem}
\newtheorem{proposition}[theorem]{Proposition}
\newtheorem{lemma}{Lemma}

\usepackage{newpxtext,newpxmath}
\usepackage[cal=boondox]{mathalfa}
\usepackage{graphicx}
\usepackage{braket}
\DeclareMathOperator{\trace}{Tr}
\NewDocumentCommand{\ketbra}{m g}{\IfNoValueTF{#2}{\ket{#1}\!\bra{#1}}{\ket{#1}\!\bra{#2}}}

\usepackage{bibunits}
\usepackage{etoolbox}
\makeatletter
\AtBeginDocument{
  \global\let\auto@bib\@empty
  \patchcmd{\endthebibliography}{\label{LastBibItem}}
    {\label{LastBibItem-\bibunitid}}{}
    {\PackageError{arxiv}{Cannot isolate bibliography labels}{}}
  \def\p@subsection{\thesection.}
}
\newcommand{\setbibunitid}[1]{
  \def\bibunitid{#1}
  \def\@extra@b@citeb{.#1}
  \def\@extra@binfo{.#1}
}
\newcommand{\clearmainbibnotes}{\let\@FMN@list\@empty}
\newcommand{\putunitbibliography}{\immediate\write\@bibunitaux
  {\string\citation{REVTEX42Control,apsrev42Control}}\expandafter\let
  \expandafter\r@LastBibItem\csname r@LastBibItem-\bibunitid\endcsname
  \putbib[bibliography,local-references,\jobname Notes]}
\makeatother
\newif\ifsupplementtoc
\let\originaladdtocontents\addtocontents
\renewcommand{\addtocontents}[2]{\ifstrequal{#1}{toc}
  {\ifsupplementtoc\originaladdtocontents{#1}{#2}\fi}
  {\originaladdtocontents{#1}{#2}}}

\usepackage[
    colorlinks=true,
    citecolor=blue,
    linkcolor=blue,
    urlcolor=blue
]{hyperref}

\begin{document}
\setbibunitid{main}
\begin{bibunit}[apsrev4-2]

\title{Extracting energy from almost unknown quantum states}

\author{Andrea Canzio}
\email{andrea.canzio@sns.it}
\affiliation{NEST, Scuola Normale Superiore, I-56126 Pisa, Italy}

\author{Vasco Cavina}
\affiliation{NEST, Scuola Normale Superiore, I-56126 Pisa, Italy}

\author{Roberto Menta}
\affiliation{NEST, Scuola Normale Superiore, I-56126 Pisa, Italy}

\author{Vittorio Giovannetti}
\affiliation{NEST, Scuola Normale Superiore, I-56126 Pisa, Italy}

\begin{abstract}
    \vspace{0pt}
    Ergotropy quantifies the maximum energy extractable by unitary operations in the state-aware, single-copy regime.
    Here we determine how much energy a fixed unitary can extract given only the mean energy.
    For any finite-dimensional Hamiltonian, the same population-reversal unitary maximizes worst-case energy extraction at every mean energy above a spectrum-dependent threshold.
    Below this threshold, no fixed unitary ensures positive extraction.
    We also characterize the minimum ergotropy at fixed mean energy and show that it coincides with optimal worst-case extraction at all energies if and only if the spectrum is antisymmetric.
    When the two differ, the gap is caused by unknown coherences: a resource for state-aware extraction becomes a liability when its structure is unknown.
    Allowing decoherence or random unitary channels removes this penalty.
    These conclusions extend to charging and any finite-dimensional observable with known mean under arbitrary unitary control, showing how prior information defines the operational value of coherence.
\end{abstract}

\maketitle
\setcounter{footnote}{0}

\textbf{\emph{Introduction.---}}Quantum thermodynamics has redefined the framework for energy manipulation at microscopic scales~\cite{e15062100,Alicki:2018dun,myers_quantum_2022,PRXQuantum.5.030309,aamir_thermally_2025,campbell2025}, opening new directions in both foundational and applied quantum science.
An extensive body of research has been developed around the maximum work extractable from and chargeable into a quantum state through unitary operations~\cite{hatsopoulos_unified_1976}, namely \emph{ergotropy}~\cite{A.E.Allahverdyan_2004} and \emph{anti-ergotropy}~\cite{salvia_distribution_2021}.
These figures of merit set fundamental limits for quantum energy conversion processes.
However, such optimal values are attainable only in the \emph{state-aware} setting.
This is a significant caveat, since in realistic settings full state tomography is often infeasible due to its exponential cost~\cite{PhysRevLett.70.1244,PhysRevA.53.4528,PhysRevA.66.012303,paris_quantum_2004,PhysRevLett.105.250403,cramer_efficient_2010,mele2024}.
Recent studies have proposed strategies to bypass this issue by relying on measurement-based protocols~\cite{vsafranek2023work,joshi2024maximal} or open quantum dynamics in the asymptotic limit~\cite{lumbreras2025quantum,watanabeUniversalWorkExtraction2026}.
Measurement-based and open systems protocols are particularly exploited in the context of quantum batteries~\cite{ferraro2018high,caravelli_energy_2021,mazzoncini2023optimal,RevModPhys.96.031001,castellano2024,Sathe_2025,Chakraborty_2025}, but can present many technical difficulties in implementation~\cite{canzio2025single}.
Non-unitary protocols also fail to capture the role of quantum coherence~\cite{ccakmak2020ergotropy,PhysRevLett.125.180603,wang2025}, entropy, and passive states~\cite{PuszWoronowicz1978,Lenard1978,bakhshinezhadTradeoffsPrecisionFluctuations2024,silvaOptimalUnitaryTrajectories2024} in energy transfer, all of which are fundamentally relevant figures of merit in the context of quantum thermodynamics.

\begin{figure}[t]
    \centering
    \includegraphics[width=\columnwidth]{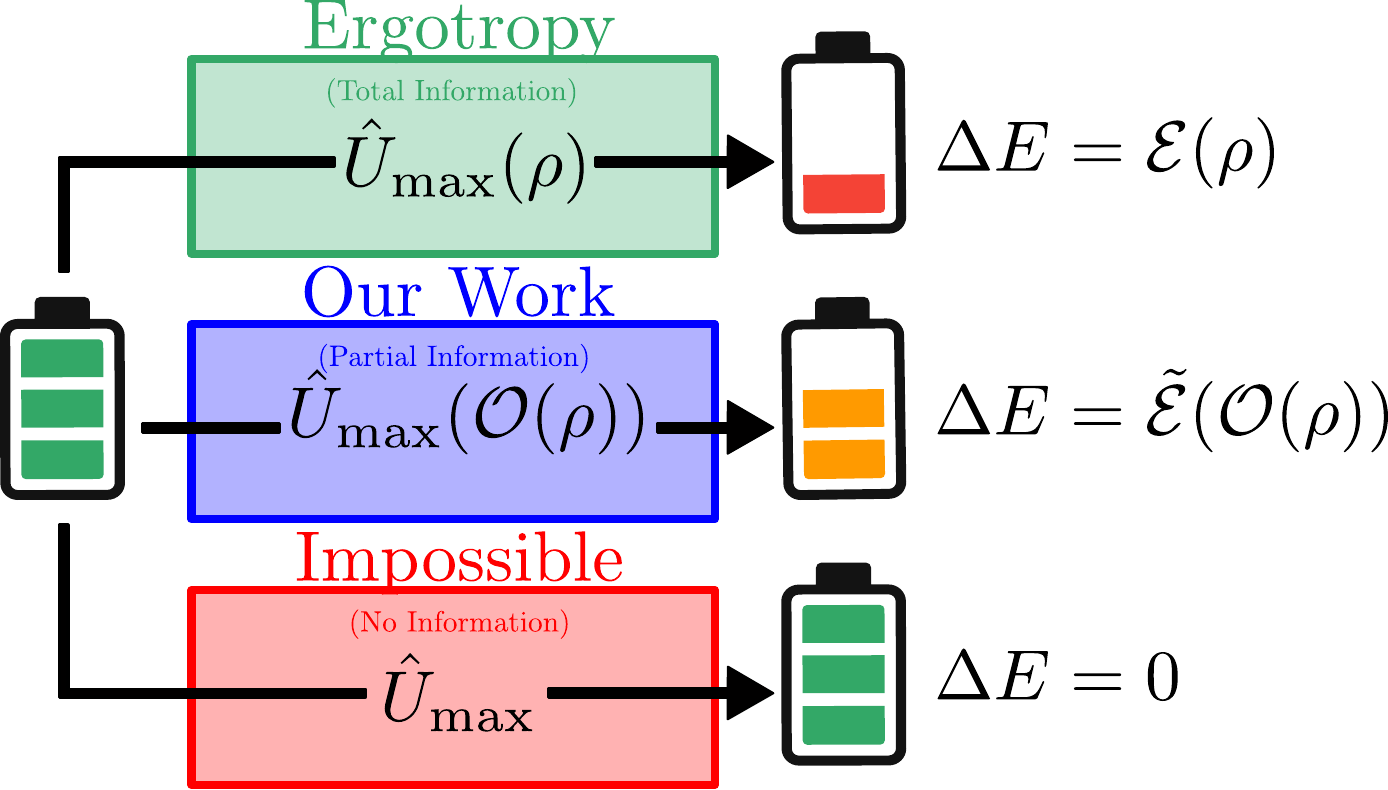}
    \caption{Schematic of the current work.
    Maximal work extraction through unitary operations (ergotropy $\mathcal{E}(\rho)$) requires full knowledge of the state $\rho$.
    Without any information about the state, no fixed unitary $\hat{U}_{\max}$ can guarantee positive energy extraction (${\Delta E>0}$).
    This work characterizes the extractable and chargeable energies when only minimal information $\mathcal{O}(\rho)$ about $\rho$ is known (specifically when $\mathcal{O}(\rho) \equiv E$).
    }
    \label{fig:intro-fig}
\end{figure}

This Letter considers a scenario in which only minimal information is already known.
Specifically, we ask:
how much energy can a fixed unitary extract from every state with mean energy
$E$? We take the Hamiltonian as known and act on a single copy without
measurements or bath coupling (Fig.~\ref{fig:intro-fig}). This setting is
motivated by batteries whose stored energy is inferred from optical
spectroscopy~\cite{quach-exp} and by systems subject to phase noise before
extraction~\cite{canzio2025single,li2025}. The same framework applies to
charging, spin ensembles with known mean
magnetization~\cite{josephDecouplingDipolarInteractions2025}, and two-mode
systems with fixed particle number and known mean population
imbalance~\cite{bardinMacroscopicQuantumSelftrapping2026}.

We start evaluating the limits of this approach.
We find upper bounds for the extractable energy in
terms of the minimum ergotropy of states with fixed mean energy.
Such states of minimum ergotropy are then characterized,
a result of independent interest,
which also quantifies the maximum values of coherent ergotropy~\cite{PhysRevLett.125.180603}.

We then go back to our original problem and solve it.
Population reversal is the optimal protocol above a
threshold set by the spectrum of the Hamiltonian.
We show that in general there is a gap between its performance
and the ergotropic upper bound.
This happens because the optimal protocol does not extract coherent energy,
and it is not optimal for incoherent energy extraction
either, unless the spectrum is antisymmetric.

Why is it the optimal one, then?
We argue that coherent energy
is actually detrimental in a state-agnostic setting,
so that the best operation you can perform is the one that
does not affect this resource.
We use this idea to reach the ergotropic bound
for any spectrum by dephasing before extraction
or averaging over suitable unitaries.
Finally, this trade-off is illustrated with a qutrit.

\textbf{\emph{Problem statement.---}}
We study optimal \emph{unitary} mean-work extraction.
Consider a $d$-dimensional quantum system with Hamiltonian $\hat{H} \coloneqq \sum_{k=1}^d \epsilon_k \ketbra{\epsilon_k}$ initially in the state $\rho$.
Its mean energy variation is
\begin{equation}
    \Delta E(\rho,\Lambda) = \trace[\rho \hat{H}] - \trace[\Lambda(\rho) \hat{H}],
\end{equation}
where $\Lambda$ is any CPTP map.

Assuming full knowledge of the system, the maximum extractable energy is quantified by its ergotropy~\cite{hatsopoulos_unified_1976,A.E.Allahverdyan_2004,salvia_distribution_2021}, $\mathcal{E}(\rho) \coloneqq \max_{\Lambda\in\mathcal{U}} \Delta E(\rho, \Lambda)$, with $\mathcal{U}$ denoting the set of all unitary channels.
The optimal protocol $\Lambda_{\max}^{(\rho)}\in\mathcal{U}$ that achieves this performance maps the eigenvectors of $\rho$ to a specific ordering of the eigenvectors of the Hamiltonian $\hat{H}$~\cite{A.E.Allahverdyan_2004} and, as such, exploits and hence requires such knowledge of the system's state.

Here, we aim to identify a protocol that enables energy extraction from any state within a given energy shell, regardless of the specific state.
Such a protocol will not attain $\mathcal{E}(\rho)$ for all states, but has the advantage of requiring much less information about the system, i.e., only its initial mean energy $E = \trace[\rho \hat{H}]$.
Formally, we want to answer the following questions: what is the largest mean-energy decrease a fixed unitary protocol can guarantee, given knowledge of only the mean energy $E$ of the state?
What protocol $\Lambda_{\max}^{(E)} \in \mathcal{U}$ achieves this performance?
We refer to the quantity that answers this question as the maximum Worst-Case Extractable Energy (max-WCEE) $\tilde{\mathcal{E}}(E)$:
\begin{equation}\label{eq:defEtilde}
    \tilde{\mathcal{E}}(E) \coloneqq \max_{\Lambda \in \mathcal{U}} \min_{\rho \in \sigma_E} \left[ \Delta E(\rho,\Lambda) \right],
\end{equation}
with $\sigma_E$ the set of all states having mean energy $E$~\footnote{Formally, $\sigma_E \coloneqq \left\{ \rho \in \sigma(\mathcal{H}) \mid \trace[\rho \hat{H}] = E \right\}$, with $\sigma(\mathcal{H})$ being the set of all states.}.

It is worth stressing the meaning of this quantity.
If $\tilde{\mathcal{E}}(E) = W$, there exists a unitary protocol $\Lambda_{\max}^{(E)}$ such that for all states with mean energy $E$ ($\rho\in\sigma_E$) it reduces the mean energy by \emph{at least} $W$~\footnote{Throughout this manuscript, we define $\arg\max_{x \in S} f(x) = \{ x_{\max} \in S \mid \forall y \in S, \; f(y)\leq f(x_{\max}) \}$.
In the SM~\cite{supplemental}, we refer at times to the same quantity as an element of the set instead of the set itself, when the specific element is not relevant.}, i.e.,
\begin{equation}
    \Delta E(\rho,\Lambda_{\max}^{(E)})\geq \tilde{\mathcal{E}}(E), \quad \forall \rho\in \sigma_E.
\end{equation}
Note that, by construction, such a protocol $\Lambda_{\max}^{(E)}$ is robust against all energy-invariant noise such as decoherence.
Observe also that max-WCEE is non-negative by construction: if no protocol can guarantee the extraction of a positive amount of energy, then by doing nothing -- i.e., applying the identity channel $\Lambda_{\mathbb{I}}$ -- we can ensure $\Delta E(\bullet,\Lambda_{\mathbb{I}}) = 0$.

We remark that our results extend beyond the specific task of work extraction.
Under the change of variables $(\hat{H}, E) \mapsto (\hat{O}, \braket{\hat{O}})$, the same mathematical framework applies to energy injection ($\hat{O} \equiv - \hat{H}$, see also Supplemental Material (SM)~\cite{supplemental}~I) and, more generally, to the minimization and maximization of any observable by means of unitary operations in the almost-agnostic regime.

\textbf{\emph{Ergotropic upper bounds.---}}
To bound what a fixed unitary can extract,
we compare its performance with what
we could extract if the state were known.
For every $\rho\in\sigma_E$,
the optimal fixed unitary extracts at least $\tilde{\mathcal{E}}(E)$
and at most $\mathcal{E}(\rho)$.
Taking the minimum over the energy shell gives
\begin{equation}\label{eq:upper}
    {\tilde{\mathcal{E}}(E)\leq\min_{\rho\in\sigma_E}\mathcal{E}(\rho) \eqqcolon\mathcal{E}_{\min}(E),}
\end{equation}
where we defined the \emph{energy-constrained minimum ergotropy} $\mathcal{E}_{\min}(E)$.
This quantity is of independent interest: it describes the least state-aware extractable work at fixed mean energy.
We now characterize it.

Passive states $\rho^\downarrow\in\Delta^\downarrow$ have zero ergotropy~\cite{hatsopoulos_unified_1976,A.E.Allahverdyan_2004,salvia_distribution_2021,PuszWoronowicz1978}.
They are diagonal in the energy eigenbasis, with populations decreasing as energy increases: $\rho^\downarrow=\sum_{k=1}^d\lambda_k^\downarrow\ketbra{\epsilon_k}$, where $\epsilon_{k+1}\geq\epsilon_k$ and $\lambda_{k+1}^\downarrow\leq\lambda_k^\downarrow$~\footnote{For degenerate $\hat{H}$, passivity and antipassivity are understood in a common eigenbasis of $\rho$ and $\hat{H}$.
The simplex and linearity statements use ordered diagonal representatives in a fixed energy eigenbasis; rotations within degenerate energy subspaces can produce additional minimizers.}.
Thus $\mathcal{E}_{\min}(E)=0$ whenever $\sigma_E$ contains a passive state.

\begin{figure*}[t]
    \centering
    \includegraphics[width=\textwidth]{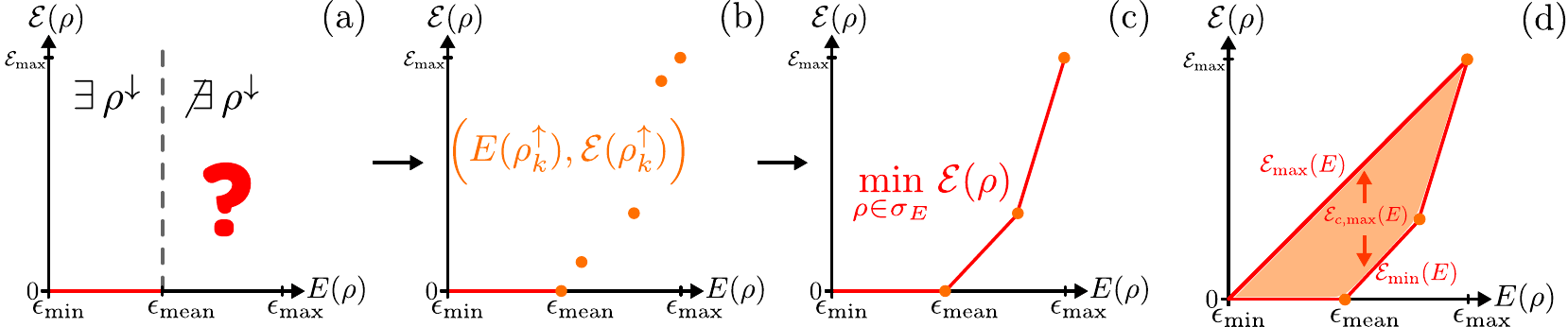}
    \caption{Energy-constrained ergotropy and its coherent contribution.
    (a) For $E\leq\epsilon_{\mathrm{mean}}$, passive states exist and the minimum ergotropy is zero.
    (b) For $E\geq\epsilon_{\mathrm{mean}}$, candidate minima are obtained from the vertices $\rho_k^\uparrow$ of the antipassive simplex $\Delta^\uparrow$ (Eq.~\eqref{eq:uppervertex}), shown as orange points in the energy--ergotropy plane.
    (c) The minimum ergotropy $\mathcal{E}_{\min}(E)$ (red line) is the lower convex envelope of these points, attained by a vertex or a convex combination of two vertices.
    (d) The allowed ergotropy values at fixed mean energy lie between $\mathcal{E}_{\min}(E)$ and $\mathcal{E}_{\max}(E)$.
    Their vertical separation gives the maximum coherent ergotropy $\mathcal{E}_{c,\max}(E)$ (Eq.~\eqref{eq:maxCErg}).}
    \label{fig:ergotropy-bounds}
\end{figure*}

However, passive states do not exist at all energies.
Let $\epsilon_{\min} \coloneqq \epsilon_1$ denote the ground-state energy, $\epsilon_{\max} \coloneqq \epsilon_d$ the maximum energy, and $\epsilon_{\mathrm{mean}} \coloneqq \frac{1}{d} \sum_{k=1}^d \epsilon_k$ the arithmetic mean of the energy levels; we have
\begin{equation}
    \exists\, \rho^\downarrow \in \sigma_E \Leftrightarrow E \in \left[ \epsilon_{\min}, \epsilon_{\mathrm{mean}} \right].
\end{equation}
Thus, we conclude that $\mathcal{E}_{\min}(E)$ (and hence $\tilde{\mathcal{E}}(E)$) is zero for mean energies $E$ that are lower than $\epsilon_{\mathrm{mean}}$, as illustrated in Fig.~\ref{fig:ergotropy-bounds}~(a).
Our first technical contribution is to characterise what happens outside this range, i.e., which states minimize ergotropy if $E\in(\epsilon_{\mathrm{mean}}, \epsilon_{\max}]$.
By defining anti-passive states $\rho^\uparrow \in \Delta^\uparrow$ analogously with eigenvalues in increasing order, we have:

\begin{proposition}\label{prop:min-states}
    Within a fixed energy shell $\sigma_E$, ergotropy is minimized by either all its passive states or some antipassive states, i.e.,
    \begin{equation}
        \begin{aligned}
            E \leq \epsilon_{\mathrm{mean}} &\Rightarrow
            \arg\min_{\rho\in\sigma_E} \mathcal{E}(\rho)
            = \sigma_E \cap \Delta^\downarrow,\\
            E \geq \epsilon_{\mathrm{mean}} &\Rightarrow
            \arg\min_{\rho\in\sigma_E} \mathcal{E}(\rho)
            \subseteq \sigma_E \cap \Delta^\uparrow.
        \end{aligned}
    \end{equation}
\end{proposition}

See SM~\cite{supplemental}~II.A for the proof.
We recall that anti-passive states are the ones that minimize the anti-ergotropy~\cite{salvia_distribution_2021} (i.e., the energy-charging equivalent of ergotropy).
Hence, Prop.~\ref{prop:min-states} tells us that above the energy threshold $\epsilon_{\rm mean}$, counterintuitively, the worst states for extraction are a subset of the worst states for charging.

For $E>\epsilon_{\mathrm{mean}}$, two observations identify the minimizing antipassive states.
First, the set of antipassive states $\Delta^\uparrow$ forms a $(d-1)$-dimensional simplex with vertices at states $\{ \rho_k^\uparrow \}_{k=1}^d$, with
\begin{equation}\label{eq:uppervertex}
    \rho_k^\uparrow \coloneqq \frac{1}{d+1-k} \sum_{i=k}^d \ketbra{\epsilon_i}.
\end{equation}
Second, the same unitary optimally discharges every state in this simplex.
When we mix the vertices, their ergotropies therefore average just as their
mean energies do. In the energy--ergotropy plane, the vertices are the orange
points in Fig.~\ref{fig:ergotropy-bounds}~(b). At a fixed energy, we look for the lowest point
we can reach by mixing the vertices. Linear programming~\cite{Boyd_Vandenberghe_2004} gives the following result (see End Matter and
SM~\cite{supplemental}~II.B).
\begin{theorem}\label{theo2}
    For $E \in [ \epsilon_{\mathrm{mean}}, \epsilon_{\max} ]$, there exists a minimum ergotropy state that is a convex combination of at most two states $\rho_k^\uparrow$ (Eq.~\eqref{eq:uppervertex}).
\end{theorem}

Consequently, $\mathcal{E}_{\min}(E)$ is piecewise linear, with breakpoints at a subset of the energies $E(\rho_k^\uparrow)$.
This piecewise-linear behavior, the algorithm to find the value of $\mathcal{E}_{\min}(E)$ and the states which attain it are illustrated in Fig.~\ref{fig:ergotropy-bounds}~(a)--(c).
Given the Hamiltonian and its (ordered) eigenvalues, the algorithm to evaluate $\mathcal{E}_{\min}(E)$ can be implemented with $O(d)$ arithmetic operations using standard linear programming techniques (see SM~\cite{supplemental}~II.C for further details).

In contrast, the maximum ergotropy at fixed mean energy admits a closed-form expression, $\mathcal{E}_{\max}(E)\coloneqq\max_{\rho\in\sigma_E}\mathcal{E}(\rho) =E-\epsilon_{\min}$, since each energy shell contains pure states that can be unitarily mapped to a ground state.
Fig.~\ref{fig:ergotropy-bounds}~(d) shows the resulting region of allowed ergotropy values.

As a consequence, this characterization also determines the maximum energy available in coherences.
The coherent ergotropy~\cite{PhysRevLett.125.180603} is defined as $\mathcal{E}_c(\rho)\coloneqq\mathcal{E}(\rho)-\mathcal{E}(\Delta(\rho))$, where $\Delta(\rho)\coloneqq\sum_k\braket{\epsilon_k|\rho|\epsilon_k} \ketbra{\epsilon_k}$ is the completely dephasing map.
Its maximum at fixed mean energy is therefore
\begin{equation}\label{eq:maxCErg}
    {\mathcal{E}_{c,\max}(E) \coloneqq\max_{\rho\in\sigma_E}\mathcal{E}_c(\rho) =\mathcal{E}_{\max}(E)-\mathcal{E}_{\min}(E).}
\end{equation}
Equality in Eq.~\eqref{eq:maxCErg} is attained by any pure state with the same populations as a diagonal minimum-ergotropy state.
Its ergotropy is maximal, whereas dephasing recovers the minimizer at the same mean energy.

\textbf{\emph{Optimal unitary extraction.---}}
We can now return to our original question.
If we know only the mean energy,
how much work can a single fixed unitary extract?
The minimum ergotropy sets an upper bound.
It is the most work we could extract from the least favourable state if we
knew it. Since this bound vanishes for $E\leq\epsilon_{\mathrm{mean}}$,
no fixed unitary can guarantee positive extraction there. How close can we
get to the bound at higher energies, and which unitary does best? For any
finite-dimensional Hamiltonian, we find that simply reversing the populations is optimal
above a spectrum-dependent threshold.

Formally, let $\hat{U}_{\mathrm{rev}}\coloneqq\sum_{k=1}^d \ketbra{\epsilon_k}{\epsilon_{d+1-k}}$ and $\Lambda_{\mathrm{rev}}(\rho)\coloneqq\hat{U}_{\mathrm{rev}}\rho\hat{U}_{\mathrm{rev}}^\dagger$ denote population reversal.
We have:

\begin{theorem}\label{th:general-unitary}
    For every $E\in[\epsilon_{\min},\epsilon_{\max}]$,
    \begin{equation}\label{eq:max-WCEE-solution}
        \tilde{\mathcal{E}}(E)=\max\!\left\{0, \min_{\rho\in\sigma_E}\Delta E(\rho,\Lambda_{\mathrm{rev}})\right\}.
    \end{equation}
    Moreover, $\tilde{\mathcal{E}}(E)>0$ if and only if $E>\epsilon_*$, where
    \begin{equation}\label{eq:energy-threshold}
        \epsilon_*\coloneqq\frac{1}{2}\max_k (\epsilon_k+\epsilon_{d+1-k})\geq\epsilon_{\mathrm{mean}}.
    \end{equation}
    Population reversal is optimal for $E\geq\epsilon_*$, while the identity (no extraction) is optimal for $E\leq\epsilon_*$.
\end{theorem}

The proof is given in SM~\cite{supplemental}~III.
Three related observations follow.
First, when $\epsilon_*>\epsilon_{\mathrm{mean}}$, all states with $\epsilon_{\mathrm{mean}}<E\leq\epsilon_*$ have positive ergotropy, yet no fixed unitary guarantees positive extraction.
The ergotropic bound is therefore strict.
Second, for a fixed Hamiltonian, the same reversal unitary works for all $E\geq\epsilon_*$; i.e., to achieve the target performance we don't need to know the specific value of $E$, but just whether $E\geq\epsilon_*$.
Third, it tells us that the optimal protocol is the one that extracts energy only from the diagonal and not the coherences, i.e., $\Delta E(\rho,\Lambda_{\mathrm{rev}}) = \Delta E(\Delta(\rho),\Lambda_{\mathrm{rev}})$.

These results show that \emph{while coherences do contain energy, without information on them they are actually detrimental.}
The remainder of the manuscript will be devoted to quantitatively proving and clarifying such a statement.

\textbf{\emph{Achievability: three special cases.---}}For generic Hamiltonians, the bound~\eqref{eq:upper} is generally not attainable.
The key insight is that while quantum coherence enhances available energy in the state-aware setting, it becomes detrimental in an almost-agnostic setting like ours.
Such an issue can be avoided by destroying coherences altogether.
This can be achieved by either restricting the analysis to diagonal states in $\sigma_E$ or by expanding the allowed operations from unitaries $\mathcal{U}$ to random unitaries $\mathcal{RU}$~\footnote{A random unitary channel $\Lambda \in \mathcal{RU}$ admits a convex decomposition into unitary channels, $\Lambda(\bullet) = \sum_k p_k \hat{U}_k \bullet \hat{U}_k^\dagger$, where $\{ p_k \}_k$ is a probability distribution and $\{\hat{U}_k \}_k \in \mathcal{U}$ are unitary operators}.
Specifically, define
\begin{align}\label{max-WCEEdiag}
    \tilde{\mathcal{E}}_{\Delta}(E)
    &\coloneqq
    \max_{\Lambda\in\mathcal{U}} \min_{\rho\in\sigma_E \cap \Delta} \left[ \Delta E(\rho,\Lambda) \right], \\
    \tilde{\mathcal{E}}_{\mathcal{RU}}(E)
    &\coloneqq
    \max_{\Lambda\in\mathcal{RU}} \min_{\rho\in\sigma_E} \left[ \Delta E(\rho,\Lambda) \right],
\end{align}
where $\Delta$ denotes the set of diagonal states in $\hat{H}$'s eigenbasis~\footnote{Here $\Delta=\{\sum_kp_k\ketbra{\epsilon_k}:p_k\geq0,\ \sum_kp_k=1\}$ in the fixed energy eigenbasis.}.
The following equivalence holds (see SM~\cite{supplemental}~IV.A for the proof):
\begin{theorem}\label{th:diag-and-rand}
    For every admissible $E$, the minimum-ergotropy bound is attained by a unitary on diagonal inputs and by a random unitary channel on arbitrary inputs.
    \begin{equation}
        \tilde{\mathcal{E}}_{\Delta}(E) = \tilde{\mathcal{E}}_{\mathcal{RU}}(E) = \mathcal{E}_{\min}(E).
    \end{equation}
\end{theorem}
In particular, observe that $\tilde{\mathcal{E}}_{\Delta}(E) = \mathcal{E}_{\min}(E)$, which is strictly greater
than $\tilde{\mathcal{E}}(E)$ over an energy interval for every non-antisymmetric spectrum.
On that interval, the population-inversion protocol $\Lambda_{\mathrm{rev}}$ is
suboptimal for diagonal inputs.
The reason is that the optimal protocols for diagonal states will generally affect coherences, and the gain from handling populations optimally is less than the loss from handling coherences incorrectly in the worst-case scenario.
Are there Hamiltonians for which the population-inversion protocol is optimal for diagonal states?
Equivalently, are there Hamiltonians for which the max-WCEE and minimum ergotropy coincide?
Eq.~\eqref{eq:energy-threshold} already points us in the right direction.

We define a Hamiltonian $\hat{H}_{\mathrm{as}} = \sum_{k=1}^d \epsilon_k \ketbra{\epsilon_k}$ to be \emph{antisymmetric} if paired eigenvalues sum to a constant: $\forall k, \epsilon_k + \epsilon_{d+1-k} = c \in \mathbb{R}$.
This property characterizes several physically relevant systems, including large spinors ($\hat{H}_{\mathrm{as}} = \hat{J}_z$) and $N$-qubit systems with arbitrary frequencies ($\hat{H}_{\mathrm{as}} = \sum_{i=1}^N \omega_i \hat{\sigma}^z_i$).
For an antisymmetric spectrum, reversal maps $\hat{H}$ to $2\epsilon_{\mathrm{mean}}\mathbb{I}_d-\hat{H}$, making the extracted energy depend only on $E$.
The following strong result follows (proofs in SM~\cite{supplemental}~IV.B):
\begin{theorem}\label{th:anti-h}
    Let $\hat{H}_{\mathrm{as}} = \sum_{k=1}^d \epsilon_k \ketbra{\epsilon_k}$ be an antisymmetric Hamiltonian.
    Then the max-WCEE is equal to the minimum ergotropy and can be written explicitly as
    \begin{equation}\label{eq:min-erg-antih}
        \tilde{\mathcal{E}}(E)
        = \mathcal{E}_{\min}(E)
        = \begin{cases}
            0 & \text{if }E \leq \epsilon_{\mathrm{mean}}, \\
            2(E-\epsilon_{\mathrm{mean}}) & \text{if } E \geq \epsilon_{\mathrm{mean}}.
        \end{cases}
    \end{equation}
\end{theorem}

We now finish by showing how coherences and populations fight for optimality with an explicit example.

\begin{figure}[t]
    \centering
    \includegraphics[width=\linewidth]{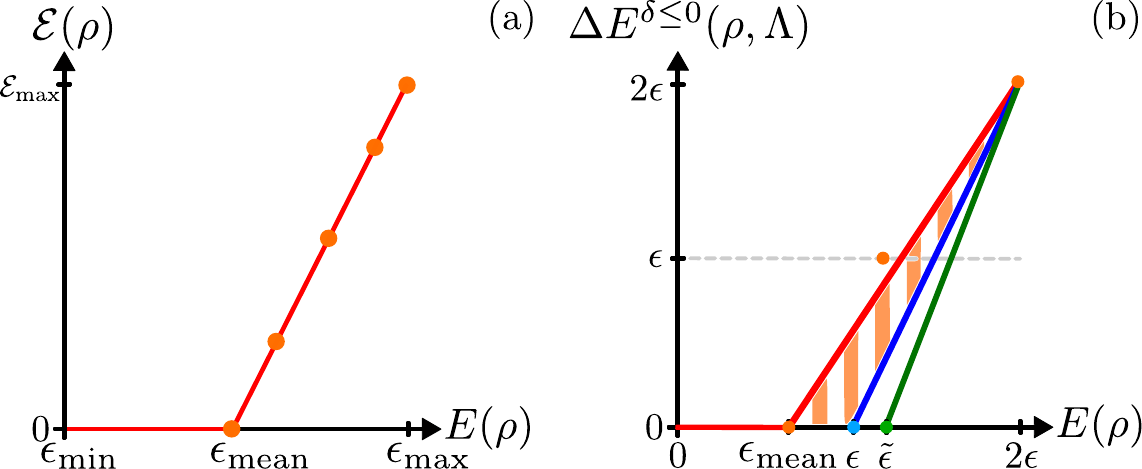}
    \caption{(a) Minimum ergotropy (solid red line) for a five-level antisymmetric Hamiltonian.
    Orange points show vertex energies and ergotropies $(E(\rho_k^\uparrow),\mathcal{E}(\rho_k^\uparrow))$ (Eq.~\eqref{eq:uppervertex}).
    All points align linearly, yielding the simple minimum ergotropy expression (Eq.~\eqref{eq:min-erg-antih}).
    $\hat{H} = \operatorname{diag}(0,0.6,2,3.4,4)$.
    (b) Comparison for a three-level system between the minimum ergotropy (red line), max-WCEE (blue line), and the minimum extractable energy through the unitary mapping $\Lambda_{\max,\Delta}^{(E)}$ that is optimal for diagonal states (green line, displayed only when positive).
    Orange stripes represent the gap between $\mathcal{E}_{\min}(E)$ and $\tilde{\mathcal{E}}(E)$.
    The orange dots are as in (a), while the light-blue dot represents $\epsilon_* = \epsilon$.
    Similarly, the light-green point $(\tilde{\epsilon},0)$, with $\tilde{\epsilon} > \epsilon$, represents the mean energy threshold at which the minimum extractable energy through $\Lambda_{\max,\Delta}^{(E)}$ becomes positive.
    Parameter: $\delta=-0.4$.
    }
    \label{fig:anti-h-and-qutrit}
\end{figure}
\begingroup
\setlength{\abovedisplayskip}{2pt}
\setlength{\belowdisplayskip}{2pt}
\setlength{\abovedisplayshortskip}{0pt}
\setlength{\belowdisplayshortskip}{2pt}
\textbf{\emph{Insights from a qutrit.---}}
Consider a qutrit with Hamiltonian $\hat{H}=\epsilon\,\operatorname{diag}(0,1+\delta,2)$, with $-1<\delta<0$.
Its non-antisymmetric spectrum produces a gap between max-WCEE and minimum ergotropy (orange stripes in Fig.~\ref{fig:anti-h-and-qutrit}~(b)).
Population reversal is therefore suboptimal for diagonal inputs.
For $E\geq\epsilon_{\mathrm{mean}}$, an optimal protocol for such inputs is (see SM~\cite{supplemental}~V)
\begin{equation}\label{eq:qutrit-diagonal-unitary}
    {
    \hat{U}_{\Delta}(\phi) =
    \begin{pmatrix}
        0 & 0 & 1 \\
        e^{i\phi} \sqrt{q} & \sqrt{1-q} & 0 \\
        e^{i\phi} \sqrt{1-q} & -\sqrt{q} & 0
    \end{pmatrix},
    }
\end{equation}
with $\phi$ arbitrary and $q = -4\delta/((1-\delta)(3-\delta))$.

With unknown coherences, its worst-case performance can fall below reversal's (Fig.~\ref{fig:anti-h-and-qutrit}~(b), green versus blue curve).
Let $\Lambda_{\phi}(\rho)=\hat{U}_{\Delta}(\phi)\rho\hat{U}_{\Delta}^\dagger(\phi)$.
Indeed, writing $\rho_{12}=\braket{\epsilon_1|\rho|\epsilon_2}$,
\begin{equation}\label{eq:qutrit-diagonal-unitary-coherence}
    { \Delta E(\rho,\Lambda_{\phi}) = \mathcal{E}_{\min}(E) + C_\delta \,\operatorname{Re} \left(e^{i\phi} \rho_{12} \right), }
\end{equation}
with $C_\delta = 2\epsilon(1-\delta)\sqrt{q(1-q)} \geq 0$.
No fixed phase can prevent a negative contribution for all coherences: given $\phi$, there always exists $\rho_{12}$ which makes the second term on the right-hand side negative.
In other words, not knowing the coherences is worse than not having them -- as promised.

Finally, let's see how random unitaries circumvent the issue.
The protocol
\begin{equation}
    \Lambda_{\mathcal{RU}} = \frac{1}{2} \left(\Lambda_{\phi} + \Lambda_{\phi+\pi} \right),
\end{equation}
averages out the coherent contribution, giving $\Delta E(\rho,\Lambda_{\mathcal{RU}})=\mathcal{E}_{\min}(E)$ for every $\rho\in\sigma_E$ and any $\phi$ (Th.~\ref{th:diag-and-rand}).
\endgroup

\penalty0\vbox{} 
\textbf{\emph{Discussion.---}}
This work addresses the question of whether, and in what manner, energy can be extracted from quantum systems through unitary operations, given knowledge of only the system’s mean energy -- an almost-agnostic and non-invasive setting.

We identify an optimal protocol for arbitrary Hamiltonians which is potentially easy to implement in experiments.
To benchmark its performance, we characterize the minimum ergotropy at fixed mean energy, which in turn quantifies the maximum coherent ergotropy -- quantities of independent interest.
The performance gap between our protocol and the benchmark reveals a fundamental quantum trade-off: known coherences improve the state-aware unitarily available energy, but unknown coherences do not, and for every Hamiltonian with a non-antisymmetric spectrum they actually reduce the worst-case extractable energy.

Interestingly, although its extracted energy depends only on the populations, the optimal protocol we find is generally not optimal for diagonal inputs.
Optimising over populations would yield a protocol which affects coherences, and the energy gain of optimal population control is lower than the energy loss of mishandling unknown coherence in the worst-case scenario.
We show this explicitly for a qutrit.

This exposes an intrinsic energy-information complementarity in quantum thermodynamic operations: the energy stored in coherence and the information needed to extract it must be considered together.
The same conclusions hold for any finite-dimensional observable under arbitrary unitary control.
A natural next question is what minimal prior information is sufficient to make coherence beneficial for guaranteed extraction.
Addressing this question may yield insights into experimentally viable protocols while deepening our understanding of the connection between energy and information in quantum systems~\cite{chattopadhyayLandauerPrincipleThermodynamics2025}.

\textbf{\emph{Acknowledgments.---}}We thank L. Lami, F. A. Mele, and R. Salvia for useful discussions.
We acknowledge financial support by MUR (Ministero dell’Universit\'a e della Ricerca) through the PNRR MUR project PE0000023-NQSTI.

\putunitbibliography

\clearpage

\section*{End Matter}

\textbf{\emph{Theorem~\ref{theo2} made simple}.}
This section aims to provide an intuitive understanding of the results in Theorem~\ref{theo2} and to characterize the states that minimize ergotropy across all mean energy values.
Prop.~\ref{prop:min-states} shows that
\begin{equation}
    \rho^* \in
    \arg\min_{\rho\in\sigma_E} \mathcal{E}(\rho)
    \Rightarrow
    \begin{cases}
        \rho^* \in \Delta^\downarrow
        & E \leq \epsilon_{\mathrm{mean}},\\
        \rho^* \in \Delta^\uparrow
        & E \geq \epsilon_{\mathrm{mean}},
    \end{cases}
\end{equation}
that is, the state that minimizes the ergotropy is passive when a passive state exists within $\sigma_E$ and is antipassive otherwise.
We note that the inequalities in Proposition~\ref{prop:min-states} are not strict because at the mean energy value $E = \epsilon_{\mathrm{mean}}$, the completely mixed state $\rho = \mathbb{I}_d/d$ minimizes the ergotropy and is both passive and antipassive.
It is well established that all passive states have zero ergotropy; therefore, for $E \leq \epsilon_{\mathrm{mean}}$, the set of ergotropy minimizers is $\sigma_E \cap \Delta^\downarrow$.
While Proposition~\ref{prop:min-states} clarifies that for $E \geq \epsilon_{\mathrm{mean}}$ the states of minimum ergotropy lie in the set $\sigma_E \cap \Delta^\uparrow$, the question remains: which specific states are they?
Theorems~\ref{theo2} and~\ref{th:anti-h} provide the precise answer.
We now outline the key observations.
\begin{itemize}
    \item The set of antipassive (and passive) states $\Delta^\uparrow$ forms a $(d-1)$-dimensional simplex.
    Consequently, there exists a fixed set of vertex states $\{ \rho_k^\uparrow \}_{k=1}^d$ such that every state $\rho^\uparrow\in\Delta^\uparrow$ can be written as a convex combination, $\rho^\uparrow = \sum_{k=1}^d p_k \rho_k^\uparrow$ (Eq.~\eqref{eq:uppervertex}).
    \item The constraint $\rho\in\sigma_E$ is a linear and convex constraint, since $\rho_1,\rho_2 \in \sigma_E$ implies that $\left(p \rho_1 + (1-p) \rho_2 \right) \in \sigma_E$ for all $p\in[0,1]$.
    \item The ergotropy is a linear functional on the set of antipassive states, since any antipassive state is transformed into the corresponding passive one by the same unitary ${\hat{U}_{\mathrm{rev}} = \sum_{k=1}^d \ketbra{\epsilon_k}{\epsilon_{d+1-k}}}$ (and similarly for anti-ergotropy on passive states).
\end{itemize}
Let us now combine these ingredients for the case $E \geq \epsilon_{\mathrm{mean}}$.
We denote the energies and ergotropies of the vertices of the antipassive state simplex as $E_k \coloneqq E(\rho_k^\uparrow)$ and $\mathcal{E}_k \coloneqq \mathcal{E}(\rho_k^\uparrow)$, respectively, and define the corresponding $d$-dimensional vectors $\boldsymbol{E}$ and $\boldsymbol{\mathcal{E}}$.
Finally, let $\boldsymbol{p} = {(p_1,\ldots,p_d)^{\mathsf{T}}}$ represent a probability vector of length $d$.
We then have:
\begin{equation}
    \begin{aligned}
        \min_{\rho\in\sigma_E} \mathcal{E}(\rho)
        &\stackrel{\text{Prop.~\ref{prop:min-states}}}{=} \min_{\rho\in \sigma_E \cap \Delta^\uparrow} \mathcal{E}(\rho),\\
        &\stackrel{\text{simplex}}{=} \min\nolimits_{\boldsymbol{p}\cdot\boldsymbol{E} = E} \mathcal{E} \left(\sum\nolimits_{k=1}^d p_k \rho_k^\uparrow \right),\\
        &\stackrel{\text{linearity}}{=} \min\nolimits_{\boldsymbol{p}\cdot\boldsymbol{E} = E} \sum\nolimits_{k=1}^d p_k \mathcal{E}(\rho_k^\uparrow),\\
        &\stackrel{\text{notation}}{=} \min\nolimits_{\boldsymbol{p}\cdot\boldsymbol{E} = E} \boldsymbol{p} \cdot \boldsymbol{\mathcal{E}},
    \end{aligned}
\end{equation}
where $\boldsymbol{a}\cdot\boldsymbol{b}$ denotes the standard scalar product of vectors $\boldsymbol{a}$ and $\boldsymbol{b}$.
This is a linear optimization problem (minimize $f(\boldsymbol{p}) = \boldsymbol{p} \cdot \boldsymbol{\mathcal{E}}$) subject to a linear constraint ($\boldsymbol{p} \cdot \boldsymbol{E} = E$).
The fundamental theorem of linear programming therefore guarantees that \emph{an} optimal solution will be attained at a vertex (i.e., a probability vector $\boldsymbol{p}$ with a single entry equal to 1) or along an edge (i.e., a vector $\boldsymbol{p}$ with exactly two non-zero entries).

\begin{figure}[t]
    \centering
    \includegraphics[width=\columnwidth]{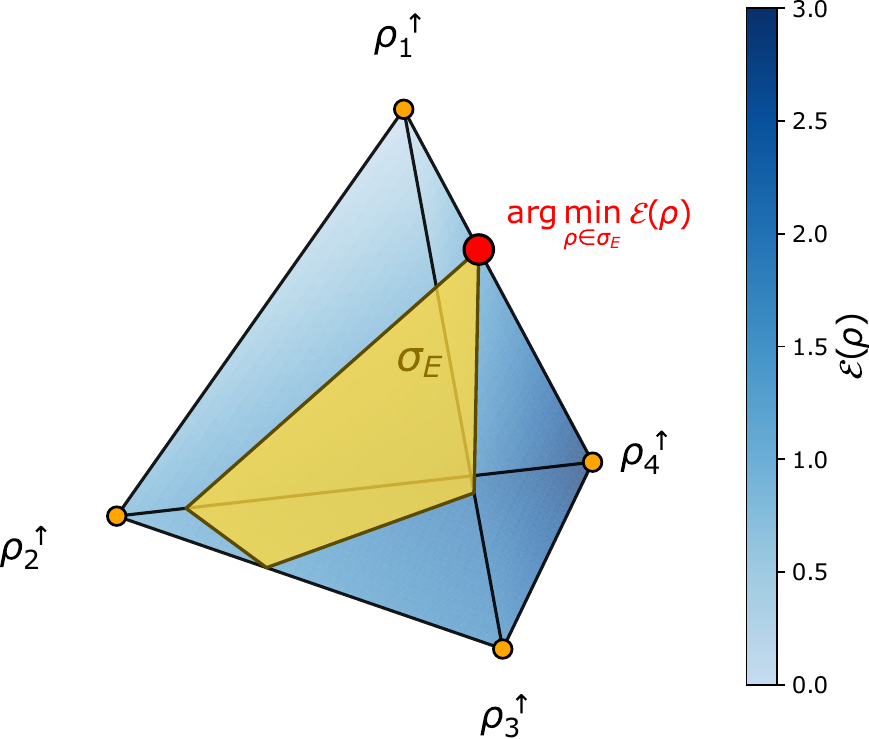}
    \caption{Graphical representation of Theorem~\ref{theo2} for a four-level system with energies $\epsilon_1 = 0$, $\epsilon_2 = 1$, $\epsilon_3 = 1.5$, $\epsilon_4 = 3$ (non-antisymmetric Hamiltonian).
    The tetrahedron represents the set $\Delta^\uparrow$ of antipassive states, with vertices $\rho_k^\uparrow$ ($k = 1, \ldots, 4$) as defined in Eq.~\eqref{eq:uppervertex}.
    The color scale indicates the ergotropy $\mathcal{E}(\rho)$, where darker blue corresponds to higher values.
    The linear constraint $\rho \in \sigma_E$ at fixed mean energy $E = 2$ is represented by the yellow plane.
    The point of minimum ergotropy on $\sigma_E \cap \Delta^\uparrow$ (red dot) lies on the edge connecting $\rho_1^\uparrow$ and $\rho_4^\uparrow$.}
    \label{fig:lin-prog}
\end{figure}
This is the statement of Theorem~\ref{theo2}, which is illustrated in Fig.~\ref{fig:lin-prog} for a four-level system: the simplex $\Delta^\uparrow$ is shown explicitly, with the ergotropy $\mathcal{E}(\rho^\uparrow) \equiv f(\boldsymbol{p})$ represented by color intensity.
The figure clearly shows that, in general, the minimum lies on an edge.
When the objective is affine in the energy on the normalized simplex, i.e., when
\begin{equation}
    {\boldsymbol{\mathcal{E}} = a\boldsymbol{E} + b\boldsymbol{1},}
\end{equation}
where $\boldsymbol{1}=(1,\ldots,1)^{\mathsf{T}}$ and $a,b\in\mathbb{R}$, the whole set $\Delta^\uparrow \cap \sigma_E$ minimizes the ergotropy.
This condition is exactly equivalent to requiring that the spectrum of the Hamiltonian is antisymmetric.
Thus, for an antisymmetric Hamiltonian and $E\geq\epsilon_{\mathrm{mean}}$, all antipassive states in $\sigma_E\cap\Delta^\uparrow$ minimize ergotropy.

\clearpage

\end{bibunit}

\onecolumngrid
\supplementtoctrue
\setbibunitid{supp}
\begin{bibunit}[apsrev4-2]
\clearmainbibnotes
\setcounter{equation}{0}
\setcounter{figure}{0}
\setcounter{section}{0}
\setcounter{subsection}{0}
\setcounter{subsubsection}{0}
\setcounter{theorem}{0}
\setcounter{lemma}{0}
\setcounter{secnumdepth}{2}
\renewcommand{\theequation}{S\arabic{equation}}
\renewcommand{\thefigure}{S\arabic{figure}}
\renewcommand{\theHequation}{supp.\arabic{equation}}
\renewcommand{\theHfigure}{supp.\arabic{figure}}
\renewcommand{\theHsection}{supp.\arabic{section}}
\renewcommand{\theHsubsection}{\theHsection.\arabic{subsection}}
\renewcommand{\theHsubsubsection}{\theHsubsection.\arabic{subsubsection}}
\providecommand{\theHtheorem}{}
\renewcommand{\theHtheorem}{supp.\arabic{theorem}}
\providecommand{\theHlemma}{}
\renewcommand{\theHlemma}{supp.\arabic{lemma}}
\begin{center}
  {\large\bfseries Supplemental Material for\\[0.4em]
  ``Extracting energy from almost unknown quantum states''}\par
  \vspace{1em}
  Andrea Canzio, Vasco Cavina, Roberto Menta, and Vittorio Giovannetti\par
  \vspace{0.4em}
  {\small NEST, Scuola Normale Superiore, I-56126 Pisa, Italy}
\end{center}
\tableofcontents
\clearpage

\section{From work extraction to charging and other tasks}
\label{sm:sec:charging}
We use the notation of the main text throughout.
In particular, $\hat{H}=\sum_{k=1}^d\epsilon_k\ketbra{\epsilon_k}$, with $\epsilon_1\leq\cdots\leq\epsilon_d$, and $\sigma_E=\{\rho\in\sigma(\mathcal{H}):\trace[\rho\hat{H}]=E\}$.
The results below allow arbitrary unitary controls, as in the main text.

The charging problem is obtained by replacing the Hamiltonian and its specified mean energy by $(-\hat{H},-E)$.
Indeed,
\begin{equation}
\label{sm:eq:charging-map}
{\mathcal{A}^{(\hat{H})}(\rho)
 \coloneqq\max_{\Lambda\in\mathcal{U}}[-\Delta E(\rho,\Lambda)]
 =\mathcal{E}^{(-\hat{H})}(\rho),\qquad
 \mathcal{A}_{\min}^{(\hat{H})}(E)
 =\mathcal{E}_{\min}^{(-\hat{H})}(-E).}
\end{equation}
We recall that $\mathcal{A}^{(\hat{H})}(\rho)$ is already well known in the literature as \emph{anti-ergotropy}~\cite{salvia_distribution_2021}.
The eigenvalues of $-\hat{H}$ in increasing order are $-\epsilon_d,\ldots,-\epsilon_1$; hence passive and antipassive states exchange roles.
Superscripts specifying the Hamiltonian will be omitted unless needed.
Every charging statement below follows from its extraction counterpart by this substitution, so no separate proofs are required.

\paragraph{Energy-constrained anti-ergotropy.}
Define the maximum worst-case injectable energy by
\begin{equation}
{\tilde{\mathcal{A}}(E)\coloneqq
 \max_{\Lambda\in\mathcal{U}}\min_{\rho\in\sigma_E}
 [-\Delta E(\rho,\Lambda)].}
\end{equation}
The converse bound and the charging version of Proposition~1 are
\begin{gather}
{0\leq\tilde{\mathcal{A}}(E)\leq\mathcal{A}_{\min}(E)
 \coloneqq\min_{\rho\in\sigma_E}\mathcal{A}(\rho),} \\
\label{sm:eq:charging-minimizers}
{\begin{aligned}
 E\geq\epsilon_{\mathrm{mean}}&\ \Rightarrow\
 \arg\min_{\rho\in\sigma_E}\mathcal{A}(\rho)
 =\sigma_E\cap\Delta^\uparrow,\\
 E\leq\epsilon_{\mathrm{mean}}&\ \Rightarrow\
 \arg\min_{\rho\in\sigma_E}\mathcal{A}(\rho)
 \subseteq\sigma_E\cap\Delta^\downarrow.
 \end{aligned}}
\end{gather}
Thus $\mathcal{A}_{\min}(E)=0$ for $E\in[\epsilon_{\mathrm{mean}},\epsilon_{\max}]$.
Here, as for Proposition~1, ordering is understood in a common eigenbasis of the state and Hamiltonian when energies are degenerate.
For $E\in[\epsilon_{\min},\epsilon_{\mathrm{mean}}]$, at least one minimizer is a convex combination of at most two passive vertices
\begin{equation}
\label{sm:eq:passive-vertices}
{\rho_k^\downarrow\coloneqq\frac{1}{k}\sum_{i=1}^k\ketbra{\epsilon_i},
 \qquad k=1,\ldots,d.}
\end{equation}
The curve is the lower convex envelope of $(E(\rho_k^\downarrow),\mathcal{A}(\rho_k^\downarrow))$.

\begin{figure}[t]
\centering
\includegraphics[width=\textwidth]{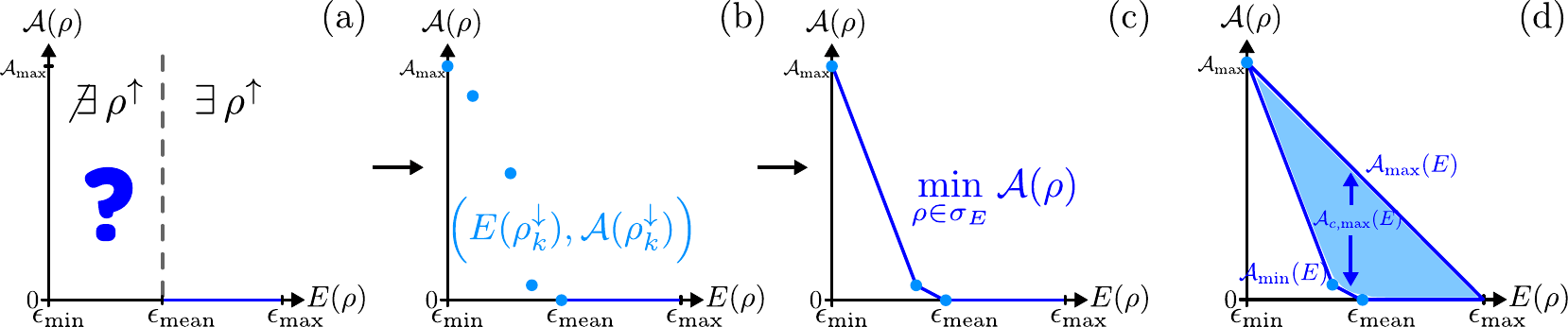}
\caption{Charging counterparts of Fig.~2 of the main text.
(a) For $E\geq\epsilon_{\mathrm{mean}}$, antipassive states exist and the minimum anti-ergotropy is zero.
(b)--(c) Below $\epsilon_{\mathrm{mean}}$, the lower convex envelope of the passive-vertex energies and anti-ergotropies (light-blue points) gives $\mathcal{A}_{\min}(E)$; a vertex or a mixture of two vertices attains every point on it.
(d) The anti-ergotropy at fixed mean energy lies between $\mathcal{A}_{\min}(E)$ and $\epsilon_{\max}-E$.
Their separation is the maximum coherent anti-ergotropy, Eq.~\eqref{sm:eq:coherent-charging}.}
\label{sm:fig:charging-bounds}
\end{figure}

\paragraph{Maximum coherent anti-ergotropy.}
In the same spirit as coherent ergotropy~\cite{PhysRevLett.125.180603}, we define the coherent anti-ergotropy by
\begin{equation}
{\mathcal{A}_c(\rho)\coloneqq\mathcal{A}(\rho)-\mathcal{A}(\Delta(\rho)),
 \qquad
 \Delta(\rho)=\sum_k\braket{\epsilon_k|\rho|\epsilon_k}\ketbra{\epsilon_k}.}
\end{equation}
The maximum anti-ergotropy at fixed mean energy is $\mathcal{A}_{\max}(E)=\epsilon_{\max}-E$, attained by pure states.
Consequently,
\begin{equation}
\label{sm:eq:coherent-charging}
{\mathcal{A}_{c,\max}(E)
 \coloneqq\max_{\rho\in\sigma_E}\mathcal{A}_c(\rho)
 =\epsilon_{\max}-E-\mathcal{A}_{\min}(E).}
\end{equation}
A pure state with the same populations as a diagonal minimum-anti-ergotropy state saturates this bound.
Fig.~\ref{sm:fig:charging-bounds} illustrates both characterizations.

\paragraph{Optimal unitary charging and achievability.}
Let $\hat{U}_{\mathrm{rev}}=\sum_k\ketbra{\epsilon_k}{\epsilon_{d+1-k}}$, $\Lambda_{\mathrm{rev}}(\rho)=\hat{U}_{\mathrm{rev}}\rho\hat{U}_{\mathrm{rev}}^\dagger$, and $\hat{H}_{\mathrm{rev}}=\hat{U}_{\mathrm{rev}}^\dagger\hat{H}\hat{U}_{\mathrm{rev}}$.
The charging counterpart of Theorem~3 is
\begin{gather}
{\tilde{\mathcal{A}}(E)=\max\{0,L_{\mathrm{rev}}(E)-E\},\qquad
 L_{\mathrm{rev}}(E)\coloneqq\min_{\rho\in\sigma_E}\trace[\rho\hat{H}_{\mathrm{rev}}],} \\
{\tilde{\mathcal{A}}(E)>0\quad\Longleftrightarrow\quad
 E<\epsilon_*^{\mathrm{ch}},\qquad
 \epsilon_*^{\mathrm{ch}}\coloneqq\frac{1}{2}\min_k
 (\epsilon_k+\epsilon_{d+1-k})\leq\epsilon_{\mathrm{mean}}.}
\end{gather}
Population reversal is optimal for $E\leq\epsilon_*^{\mathrm{ch}}$, and the identity is optimal for $E\geq\epsilon_*^{\mathrm{ch}}$.
Defining $\tilde{\mathcal{A}}_{\Delta}$ and $\tilde{\mathcal{A}}_{\mathcal{RU}}$ by restricting the input states or allowing random unitary channels, respectively, as in Theorem~4, gives
\begin{equation}
{\tilde{\mathcal{A}}_{\Delta}(E)
 =\tilde{\mathcal{A}}_{\mathcal{RU}}(E)=\mathcal{A}_{\min}(E).}
\end{equation}
Finally, for an antisymmetric spectrum, Theorem~5 becomes
\begin{equation}
{\tilde{\mathcal{A}}(E)=\mathcal{A}_{\min}(E)
 =\max\{0,2(\epsilon_{\mathrm{mean}}-E)\}.}
\end{equation}

\paragraph{Other observables.}
The same framework applies to any finite-dimensional Hermitian observable $\hat{O}$ whose initial mean $o$ is known: replace $(\hat{H},E)$ by $(\hat{O},o)$ everywhere.
The decrease in its mean is $\trace[\rho\hat{O}]-\trace[\Lambda(\rho)\hat{O}]$; increases follow by replacing $\hat{O}$ by $-\hat{O}$.
Examples include magnetization or particle number, provided the same class of unconstrained unitary controls is available.
Additional conservation laws or superselection restrictions on the controls define a different optimization problem.

\section{Energy-constrained minimum ergotropy}
\label{sm:sec:minimum}

\subsection{Proof of Proposition~1}
\label{sm:sec:prop1}
\begin{proposition}
\label{sm:prop:min-states}
Within a fixed energy shell $\sigma_E$, ergotropy is minimized by either all its passive states or some antipassive states, i.e.,
\begin{equation}
\begin{aligned}
            E \leq \epsilon_{\mathrm{mean}} &\Rightarrow
        \arg\min_{\rho\in\sigma_E} \mathcal{E}(\rho)
        = \sigma_E \cap \Delta^\downarrow,\\
        E \geq \epsilon_{\mathrm{mean}} &\Rightarrow
        \arg\min_{\rho\in\sigma_E} \mathcal{E}(\rho)
        \subseteq \sigma_E \cap \Delta^\uparrow.
        \end{aligned}
\end{equation}
\end{proposition}
We first recall the spectral expression for ergotropy.
Let $\rho=\sum_i\lambda_i\ketbra{\phi_i}$ be a generic state with mean energy $E$.
Its ergotropy can be written as~\cite{salvia_distribution_2021}
\begin{equation}
\label{sm:eq:spectral-ergotropy}
 \mathcal{E}(\rho)=E-E_-(\rho),\qquad
 E_-(\rho)\coloneqq\sum_{k=1}^d\lambda_k^\downarrow\epsilon_k,
 \qquad
 E_+(\rho)\coloneqq\sum_{k=1}^d\lambda_k^\uparrow\epsilon_k,
\end{equation}
where $\lambda_k^\downarrow$ ($\lambda_k^\uparrow$) are the eigenvalues of $\rho$ in decreasing (increasing) order.
The quantities $E_-$ and $E_+$ are the minimum and maximum energies on its unitary orbit.
For degenerate $\hat{H}$, passivity and antipassivity are understood in a common eigenbasis of $\rho$ and $\hat{H}$; statements about a simplex refer to ordered diagonal representatives in one fixed energy eigenbasis.

\subsubsection{Reduction to diagonal representatives}
We show that from every $\rho\in\sigma_E$ we can construct another $\tilde{\rho}\in\sigma_E\cap\Delta$ such that $\mathcal{E}(\tilde{\rho})\leq\mathcal{E}(\rho)$.
Consider the energy-preserving quantum channel that kills the coherences with respect to a fixed Hamiltonian eigenbasis:
\begin{equation}
\tilde{\rho}=\Delta(\rho)
 =\sum_{i=1}^d\braket{\epsilon_i|\rho|\epsilon_i}\ketbra{\epsilon_i}
 =\sum_{i,j=1}^d\lambda_j|\braket{\phi_j|\epsilon_i}|^2\ketbra{\epsilon_i}
 =\sum_i\gamma_i\ketbra{\epsilon_i}.
\end{equation}
Here $\boldsymbol{\gamma}=M\boldsymbol{\lambda}$, with $M_{ij}=|\braket{\phi_j|\epsilon_i}|^2$.
The matrix $M$ is doubly stochastic; thus, we can write it as a convex sum of permutation matrices, $M=\sum_{\pi} p_{\pi} P_{\pi}$, where $\sum_{\pi} p_{\pi}=1$~\cite{DUFOSSE2016108}.
Let $P_{\gamma}^\downarrow$ order $\boldsymbol{\gamma}$ decreasingly.
Then
\begin{equation}
\begin{aligned}
\mathcal{E}(\rho)-\mathcal{E}(\tilde{\rho})
 &=\boldsymbol{\gamma}^\downarrow\cdot\boldsymbol{\epsilon}
   -\boldsymbol{\lambda}^\downarrow\cdot\boldsymbol{\epsilon}\\
 &=\sum_{\pi} p_{\pi}
 (P_{\gamma}^\downarrow P_{\pi}\boldsymbol{\lambda})\cdot\boldsymbol{\epsilon}
   -\boldsymbol{\lambda}^\downarrow\cdot\boldsymbol{\epsilon}\geq0.
\end{aligned}
\end{equation}
The first term is a convex combination of scalar products between permutations of the original spectrum and the ordered energy levels.
By construction, $\boldsymbol{\lambda}^\downarrow\cdot\boldsymbol{\epsilon}$ is already the minimum over these permutations.
We have therefore shown that the minimum over $\sigma_E$ equals the minimum over $\sigma_E\cap\Delta$.
This proves the existence of a diagonal minimizer; it does not by itself characterize all minimizers.

\subsubsection{Passive and antipassive minimizers}
\begin{proof}[Proof of Proposition~\ref{sm:prop:min-states}]
All passive states have zero ergotropy.
Up to a unitary commuting with $\hat{H}$, write such a state in the fixed energy eigenbasis, with populations ordered as $\lambda_1\geq\cdots\geq\lambda_d$. Set $\lambda_{d+1}=0$ and $p_k=k(\lambda_k-\lambda_{k+1})$.
Then $p_k\geq0$, $\sum_kp_k=1$, and $\rho^\downarrow=\sum_kp_k\rho_k^\downarrow$, with $\rho_k^\downarrow$ defined in Eq.~\eqref{sm:eq:passive-vertices}.
This within-eigenspace change of basis preserves both energy and ergotropy.
Every passive state therefore has mean energy at most $\epsilon_{\mathrm{mean}}$.
Conversely, mixtures of a ground state and $\mathbb{I}_d/d$ give a passive state at every energy in $[\epsilon_{\min},\epsilon_{\mathrm{mean}}]$.
Since zero ergotropy is equivalent to passivity, the first equality follows.

For $E>\epsilon_{\mathrm{mean}}$, minimizing ergotropy corresponds to maximizing $E_-(\rho)$ at fixed $E$.
We show directly that every minimizer is antipassive.
Suppose that a minimizing state $\rho$ has $E_+(\rho)>E$.
Let $\rho^\uparrow$ be an antipassive state on its unitary orbit, so $\trace[\rho^\uparrow\hat{H}]=E_+(\rho)$.
Consider
\begin{equation}
{\tau=t\rho^\uparrow+(1-t)\frac{\mathbb{I}_d}{d},\qquad
 t=\frac{E-\epsilon_{\mathrm{mean}}}{E_+(\rho)-\epsilon_{\mathrm{mean}}}\in(0,1).}
\end{equation}
By construction, $\tau$ is antipassive and belongs to $\sigma_E$.
Its eigenvalues are $t\lambda_i+(1-t)/d$, with the same ordering as those of $\rho$, and hence
\begin{equation}
{E_-(\tau)=tE_-(\rho)+(1-t)\epsilon_{\mathrm{mean}}>E_-(\rho).}
\end{equation}
The inequality is strict for nonconstant $\hat{H}$ and nonuniform $\rho$: oppositely ordered spectra obey
\begin{equation}
{E_-(\rho)-\epsilon_{\mathrm{mean}}
 =\frac{1}{d}\sum_{i<j}
 (\lambda_i^\downarrow-\lambda_j^\downarrow)(\epsilon_i-\epsilon_j)<0.}
\end{equation}
At least the pair of extreme eigenvalues gives a strict term.
The state $\rho$ cannot be uniform because its mean energy exceeds $\epsilon_{\mathrm{mean}}$.
Thus $\mathcal{E}(\tau)<\mathcal{E}(\rho)$, contradicting minimality.
Every minimizer must instead satisfy $E_+(\rho)=E$, which is exactly the condition of antipassivity.
Indeed, stationarity of the orbit energy gives $[\rho,\hat{H}]=0$, and exchanging two common eigenvectors excludes decreasing populations across strictly increasing energies at an orbit maximum.

At $E=\epsilon_{\mathrm{mean}}$, a minimizer has zero ergotropy and is passive.
For nonconstant $\hat{H}$, the same strict ordering identity shows that its populations must all equal $1/d$, so it is also antipassive.
If $\hat{H}$ is proportional to the identity, every state is both passive and antipassive and the proposition is immediate.
\end{proof}

\subsection{Proof of Theorem~2}
\label{sm:sec:th2}
Recall the antipassive vertices of the main text,
\begin{equation}
\label{sm:eq:uppervertex}
 \rho_k^\uparrow\coloneqq\frac{1}{d+1-k}
 \sum_{i=k}^d\ketbra{\epsilon_i},\qquad k=1,\ldots,d.
\end{equation}
\begin{theorem}
\label{sm:theo2}
For $E \in [ \epsilon_{\mathrm{mean}} , \epsilon_{\max} ]$, there exists a minimum ergotropy state that is a convex combination of at most two states $\rho_k^\uparrow$ (Eq.~\eqref{sm:eq:uppervertex}).
\end{theorem}
\begin{proof}
Thanks to Proposition~\ref{sm:prop:min-states}, the energy-constrained minimum ergotropy is achieved by an antipassive state for $E\geq\epsilon_{\mathrm{mean}}$.
A unitary commuting with $\hat{H}$ can map this state to its ordered diagonal representative in the fixed energy eigenbasis, without changing its energy or ergotropy.
In order to characterize such a state, we observe that the ordered diagonal representatives form a $(d-1)$-dimensional simplex.
Indeed, if $\rho^\uparrow=\sum_k\lambda_k^\uparrow\ketbra{\epsilon_k}$, set $\lambda_0^\uparrow=0$ and
\begin{equation}
p_k=(d+1-k)(\lambda_k^\uparrow-\lambda_{k-1}^\uparrow).
\end{equation}
Clearly $p_k\geq0$, and telescoping gives
\begin{equation}
\sum_kp_k=\sum_k\lambda_k^\uparrow=1,\qquad
 \rho^\uparrow=\sum_kp_k\rho_k^\uparrow.
\end{equation}
The differences of successive populations fix the $p_k$ uniquely, so these states are the vertices of a simplex.

We now recall that ergotropy is linear on this simplex.
This happens because all states in this ordered simplex are optimally discharged by the same unitary $\hat{U}_{\mathrm{rev}}$.
It follows that the minimization problem can be rewritten as
\begin{equation}
\label{sm:eq:minimum-lp}
 \mathcal{E}_{\min}(E)
 =\min_{\substack{p_k\geq0,\ \sum_kp_k=1\\
                   \sum_kp_kE_k=E}}\sum_kp_k\mathcal{E}_k,
 \qquad E_k\coloneqq\trace[\hat{H}\rho_k^\uparrow],\quad
 \mathcal{E}_k\coloneqq\mathcal{E}(\rho_k^\uparrow).
\end{equation}
This is a linear program, and we now want to show that its minimum can always be achieved by either a vertex or a convex combination of two vertices~\cite{Boyd_Vandenberghe_2004}.
Let $\boldsymbol{p}^*$ be an optimal solution with more than two positive entries.
If the supported energies are not all equal, select three supported indices including a pair with distinct energies, $E_a\ne E_b$, and denote the third by $h$.
Define a vector $\boldsymbol{t}$ supported on these three indices by
\begin{equation}
t_h=-1,\qquad t_a=\frac{E_b-E_h}{E_b-E_a},\qquad
 t_b=1-t_a.
\end{equation}
Then $\sum_kt_k=0$ and $\sum_kt_kE_k=0$.
For sufficiently small positive or negative $\eta$, $\boldsymbol{p}^*+\eta\boldsymbol{t}$ is a feasible probability distribution.
Optimality implies
\begin{equation}
(\boldsymbol{p}^*+\eta\boldsymbol{t})\cdot\boldsymbol{\mathcal{E}}
 \geq\boldsymbol{p}^*\cdot\boldsymbol{\mathcal{E}}.
\end{equation}
Since $\eta$ can have either sign, we must have $\boldsymbol{t}\cdot\boldsymbol{\mathcal{E}}=0$.
Moving along this direction until an entry vanishes preserves optimality and reduces the support.
If all supported vertex energies coincide, transfer probability between any two supported indices instead; the same argument applies.
Repeating these steps leaves at most two positive entries.
This proves existence, without requiring uniqueness of the minimizer.
\end{proof}

\subsection{An algorithm for the energy-constrained minimum ergotropy and its complexity}
\label{sm:sec:algorithm}
As shown in the previous subsection, finding the minimum-ergotropy curve amounts to evaluating the lower convex envelope of the points $(E_k,\mathcal{E}_k)$.
A single scan suffices because the vertex energies are already ordered.
For $m=d+1-k$, write
\begin{equation}
\label{sm:eq:vertex-coordinates}
{X_m\coloneqq\frac{1}{m}\sum_{i=d-m+1}^d\epsilon_i,
 \qquad
 Y_m\coloneqq X_m-\frac{1}{m}\sum_{i=1}^m\epsilon_i,
 \qquad (E_k,\mathcal{E}_k)=(X_m,Y_m).}
\end{equation}
Prefix and suffix sums give all $d$ pairs in $O(d)$ arithmetic operations.
Processing $m=d,d-1,\ldots,1$ puts them in nondecreasing order of $X_m$, since $X_{m-1}-X_m=(X_m-\epsilon_{d-m+1})/(m-1)\geq0$ for $m\geq2$.
The following procedure constructs the required lower boundary.
\begin{enumerate}
\item
 Generate these pairs in that order, retaining their vertex indices.
For repeated abscissas, retain a pair with the smallest ordinate.
\item
 Maintain a list $L$, initially empty.
Before appending a new point $C$, let $A,B$ be its last two points, if present.
While
\begin{equation}
\label{sm:eq:hull-turn}
{(X_B-X_A)(Y_C-Y_A)-(Y_B-Y_A)(X_C-X_A)\leq0,}
\end{equation}
remove $B$ and repeat the test with the new last two points.
Then append $C$.
\item
 Connect successive retained points by straight segments.
This is $\mathcal{E}_{\min}(E)$ on $[\epsilon_{\mathrm{mean}},\epsilon_{\max}]$; extend it by zero on $[\epsilon_{\min},\epsilon_{\mathrm{mean}}]$.
\end{enumerate}
Each point is appended once and removed at most once, so the scan uses $O(d)$ operations and $O(d)$ memory.
If $A,B$ bracket a requested energy,
\begin{equation}
\label{sm:eq:hull-state}
{\mathcal{E}_{\min}(E)=\frac{X_B-E}{X_B-X_A}Y_A
 +\frac{E-X_A}{X_B-X_A}Y_B,\qquad
 \rho_E=\frac{X_B-E}{X_B-X_A}\rho_A^\uparrow
 +\frac{E-X_A}{X_B-X_A}\rho_B^\uparrow}
\end{equation}
gives both the minimum and one state attaining it; at a retained point, use its vertex state directly.
Here $\rho_A^\uparrow,\rho_B^\uparrow$ denote the vertex states whose indices were stored in step~1.
If only one distinct abscissa remains, the high-energy interval is a single point.
The non-strict test removes redundant collinear points, which does not change the curve; keeping them records additional vertex minimizers on the same face.

The $O(d)$ bound assumes that the ordered eigenvalues are supplied, as in the main text, and counts real-arithmetic operations.
Sorting an unordered spectrum adds $O(d\log d)$ operations; obtaining the spectrum from a matrix is a separate task.
Once the boundary is stored, a single query takes $O(\log h)$ time by binary search among its $h\leq d$ vertices.
Thus repeated slope searches, which can take $O(d^2)$ operations, are unnecessary.
The construction solves the linear program \eqref{sm:eq:minimum-lp} simultaneously for every energy.

\subsection{Maximum energy-constrained coherent ergotropy}
\label{sm:sec:coherent}
A consequence of the characterization of the energy-constrained minimum ergotropy is an optimal upper bound for coherent ergotropy at fixed mean energy.
We recall that coherent ergotropy~\cite{PhysRevLett.125.180603} is defined as the difference between the ergotropy of the state and that of its dephased version,
\begin{equation}
\mathcal{E}_c(\rho)\coloneqq\mathcal{E}(\rho)-\mathcal{E}(\Delta(\rho)).
\end{equation}
As stated in the main text,
\begin{equation}
\label{sm:eq:maxCErg}
 \mathcal{E}_{c,\max}(E)\coloneqq\max_{\rho\in\sigma_E}\mathcal{E}_c(\rho)
 =\mathcal{E}_{\max}(E)-\mathcal{E}_{\min}(E),\qquad
 \mathcal{E}_{\max}(E)=E-\epsilon_{\min}.
\end{equation}
\begin{proof}
Thanks to the reduction to diagonal representatives in Sec.~\ref{sm:sec:prop1}, we have
\begin{equation}
\begin{aligned}
\mathcal{E}_{c,\max}(E)
 &=\max_{\rho\in\sigma_E}
 [\mathcal{E}(\rho)-\mathcal{E}(\Delta(\rho))]\\
 &\leq\max_{\rho\in\sigma_E}\mathcal{E}(\rho)
       -\min_{\rho\in\sigma_E}\mathcal{E}(\Delta(\rho))
 =E-\epsilon_{\min}-\mathcal{E}_{\min}(E).
\end{aligned}
\end{equation}
In the last step we exploited the fact that a minimizing state can be chosen diagonal, so minimizing over dephased states gives the same value as minimizing over all states.
We also used $\max_{\rho\in\sigma_E}\mathcal{E}(\rho)=E-\epsilon_{\min}$, since pure states of every admissible mean energy exist and can be mapped to a ground state.
Now let $\tilde{\rho}=\sum_kp_k\ketbra{\epsilon_k}\in\sigma_E$ be a diagonal minimum-ergotropy state.
The inequality is saturated by the pure state
\begin{equation}
\ket{\Psi}=\sum_k\sqrt{p_k}e^{i\theta_k}\ket{\epsilon_k},
\end{equation}
for any phases $\theta_k$: its ergotropy is $E-\epsilon_{\min}$, while $\Delta(\ketbra{\Psi})=\tilde{\rho}$.
\end{proof}

\section{Maximum worst-case extractable energy (max-WCEE)}
\label{sm:sec:maxwcee}

\subsection{Preliminaries}
\label{sm:sec:unitary-preliminaries}
We prove the two lemmas needed to determine the optimal unitary protocol.
They hold for every finite-dimensional Hamiltonian, including degenerate spectra and the endpoint energy shells.
The first lemma expresses the maximization over states as a scalar infimum.
The second evaluates the resulting minimization over unitaries.
Together, they give the proof of Theorem~3.

\begin{lemma}[Scalar dual of the energy constraint]
\label{sm:lem:dual}
For every Hermitian operator $A$ and every admissible mean energy $E$,
\begin{equation}
\label{sm:eq:scalar-dual}
{\max_{\rho\in\sigma_E}\trace[\rho A]
 =\inf_{\beta\in\mathbb{R}}
 \{\beta E+\lambda_{\max}(A-\beta\hat{H})\}.}
\end{equation}
\end{lemma}
\begin{proof}
For every $\rho\in\sigma_E$ and $\beta\in\mathbb{R}$,
\begin{equation}
{\trace[\rho A]=\beta E+\trace[\rho(A-\beta\hat{H})]
 \leq\beta E+\lambda_{\max}(A-\beta\hat{H}).}
\end{equation}
This proves one inequality.
For the opposite inequality, consider the compact convex set
\begin{equation}
{K_A=\{(\trace[\rho\hat{H}],\trace[\rho A]):\rho\in\sigma(\mathcal{H})\}.}
\end{equation}
Its upper boundary is the concave function $f_A(x)=\max_{\rho\in\sigma_x}\trace[\rho A]$.
For $\epsilon_{\min}<E<\epsilon_{\max}$, it admits an upper supporting line $f_A(x)\leq f_A(E)+\beta(x-E)$ for some finite $\beta$.
Maximizing over all states gives
\begin{equation}
{\lambda_{\max}(A-\beta\hat{H})
 =\max_x[f_A(x)-\beta x]=f_A(E)-\beta E,}
\end{equation}
which proves equality.

At the endpoints a finite supporting slope need not exist, which is why Eq.~\eqref{sm:eq:scalar-dual} uses an infimum.
For $E=\epsilon_{\min}$ let $P$ project onto the ground-energy subspace and set $B=\hat{H}-\epsilon_{\min}\mathbb{I}_d\geq0$.
Every admissible state is supported on $\operatorname{ran}P$, so $f_A(\epsilon_{\min})=\lambda_{\max}(PAP|_{\operatorname{ran}P})$.
The dual expression is then $g(\beta)=\lambda_{\max}(A-\beta B)$.
For a normalized maximizing eigenvector $v_\beta$, comparison with a vector in $\ker B$ yields, for $\beta>0$,
\begin{equation}
{g(\beta)\geq f_A(\epsilon_{\min}),\qquad
 0\leq\beta\braket{v_\beta|B|v_\beta}
 \leq\|A\|-f_A(\epsilon_{\min}).}
\end{equation}
Every accumulation point of $v_\beta$ as $\beta\to+\infty$ therefore lies in $\ker B$, while $g(\beta)\leq\braket{v_\beta|A|v_\beta}$.
Thus $g(\beta)\to f_A(\epsilon_{\min})$.
The upper endpoint follows by the same argument with $B=\epsilon_{\max}\mathbb{I}_d-\hat{H}$ and $\beta\to-\infty$.
If $\hat{H}$ is scalar, Eq.~\eqref{sm:eq:scalar-dual} follows directly.
\end{proof}

\begin{lemma}[Spectral alignment]
\label{sm:lem:alignment}
For every real $\beta$,
\begin{equation}
\label{sm:eq:spectral-alignment}
{\min_{\hat{U}}\lambda_{\max}(\hat{U}^\dagger\hat{H}\hat{U}-\beta\hat{H})
 =\begin{cases}
 \max_k(1-\beta)\epsilon_k,&\beta\geq0,\\
 \max_k(\epsilon_{d+1-k}-\beta\epsilon_k),&\beta\leq0.
 \end{cases}}
\end{equation}
The minima are attained by the identity and population reversal, respectively.
\end{lemma}
\begin{proof}
Let $t=\lambda_{\max}(\hat{U}^\dagger\hat{H}\hat{U}-\beta\hat{H})$.
Then
\begin{equation}
{\hat{U}^\dagger\hat{H}\hat{U}\leq t\mathbb{I}_d+\beta\hat{H}.}
\end{equation}
Operator order implies the same order for increasingly ordered eigenvalues: $B\leq C$ gives $\lambda_k^\uparrow(B)\leq\lambda_k^\uparrow(C)$ by the min--max principle, since every Rayleigh quotient of $B$ is at most that of $C$.
For $\beta\geq0$, the right-hand eigenvalues are $t+\beta\epsilon_k$ in increasing order.
Hence $\epsilon_k\leq t+\beta\epsilon_k$, and $t\geq\max_k(1-\beta)\epsilon_k$.
The identity attains this bound.
For $\beta\leq0$, the ordering reverses and gives $\epsilon_k\leq t+\beta\epsilon_{d+1-k}$, so
\begin{equation}
{t\geq\max_k(\epsilon_k-\beta\epsilon_{d+1-k})
 =\max_k(\epsilon_{d+1-k}-\beta\epsilon_k).}
\end{equation}
Population reversal attains the bound.
The argument includes degeneracies.
\end{proof}

\subsection{Proof of Theorem~3}
\label{sm:sec:th3}
\begin{theorem}[Optimal unitary extraction]
\label{sm:th:general-unitary}
For every $E\in[\epsilon_{\min},\epsilon_{\max}]$,
\begin{equation}
\label{sm:eq:max-WCEE-solution}
{\tilde{\mathcal{E}}(E)=\max\!\left\{0,
\min_{\rho\in\sigma_E}\Delta E(\rho,\Lambda_{\mathrm{rev}})\right\}.}
\end{equation}
Moreover, $\tilde{\mathcal{E}}(E)>0$ if and only if $E>\epsilon_*$, where
\begin{equation}
\label{sm:eq:energy-threshold}
{\epsilon_*\coloneqq\frac{1}{2}\max_k
(\epsilon_k+\epsilon_{d+1-k})\geq\epsilon_{\mathrm{mean}}.}
\end{equation}
Population reversal is optimal for $E\geq\epsilon_*$, while the identity (no extraction) is optimal for $E\leq\epsilon_*$.
\end{theorem}
\begin{proof}
It is convenient to minimize the worst-case final energy.
Define
\begin{equation}
{A_U=\hat{U}^\dagger\hat{H}\hat{U},\qquad
 F_U(E)=\max_{\rho\in\sigma_E}\trace[\rho A_U],\qquad
 F_*(E)=\min_{\hat{U}}F_U(E).}
\end{equation}
Then $\tilde{\mathcal{E}}(E)=E-F_*(E)$.
Applying Lemma~\ref{sm:lem:dual} and exchanging two infima gives
\begin{equation}
\label{sm:eq:two-infima}
{\begin{aligned}
F_*(E)
 &=\inf_{\hat{U}}\inf_{\beta\in\mathbb{R}}
 \{\beta E+\lambda_{\max}(\hat{U}^\dagger\hat{H}\hat{U}-\beta\hat{H})\}\\
 &=\inf_{\beta\in\mathbb{R}}\left\{\beta E+
 \min_{\hat{U}}\lambda_{\max}(\hat{U}^\dagger\hat{H}\hat{U}-\beta\hat{H})\right\}.
\end{aligned}}
\end{equation}
This exchanges two minimizations over $(\hat{U},\beta)$, rather than interchanging the maximization over protocols with the minimization over states.
Convexity of the unitary group is not required.

Using Lemma~\ref{sm:lem:alignment}, the branch $\beta\geq0$ gives
\begin{equation}
{\inf_{\beta\geq0}\{\beta E+\max_k(1-\beta)\epsilon_k\}=E.}
\end{equation}
Indeed, for $0\leq\beta\leq1$ its integrand is $E+(1-\beta)(\epsilon_{\max}-E)\geq E$; for $\beta\geq1$ it is $E+(\beta-1)(E-\epsilon_{\min})\geq E$.
Equality holds at $\beta=1$.

For the branch $\beta\leq0$, set
\begin{equation}
\label{sm:eq:Rrev}
{R_{\mathrm{rev}}(E)\coloneqq\max_{\rho\in\sigma_E}
 \trace[\rho\hat{H}_{\mathrm{rev}}],\qquad
 \hat{H}_{\mathrm{rev}}=\hat{U}_{\mathrm{rev}}^\dagger\hat{H}\hat{U}_{\mathrm{rev}}
 =\sum_k\epsilon_{d+1-k}\ketbra{\epsilon_k}.}
\end{equation}
Lemma~\ref{sm:lem:dual} expresses $R_{\mathrm{rev}}(E)$ as the infimum of $\beta E+\max_k(\epsilon_{d+1-k}-\beta\epsilon_k)$ over all real $\beta$.
Restricting to $\beta\leq0$ does not change this value: for $\beta\geq0$,
\begin{equation}
{\beta E+\max_k(\epsilon_{d+1-k}-\beta\epsilon_k)
 =\epsilon_{\max}+\beta(E-\epsilon_{\min})\geq\epsilon_{\max},}
\end{equation}
and the value $\epsilon_{\max}$ is already available at $\beta=0$.
Consequently,
\begin{equation}
{F_*(E)=\min\{E,R_{\mathrm{rev}}(E)\},\qquad
 \tilde{\mathcal{E}}(E)=\max\{0,E-R_{\mathrm{rev}}(E)\}.}
\end{equation}
The identity and reversal realize these two alternatives, proving the claimed operational optimum.

It remains to determine the threshold.
For any state in $\sigma_E$, writing its populations as $p_k$ gives
\begin{equation}
{E+\trace[\rho\hat{H}_{\mathrm{rev}}]
 =\sum_kp_k(\epsilon_k+\epsilon_{d+1-k})\leq2\epsilon_*.}
\end{equation}
Hence $E-R_{\mathrm{rev}}(E)\geq2(E-\epsilon_*)>0$ when $E>\epsilon_*$.
Conversely, choose $k$ maximizing the paired sum and consider
\begin{equation}
{\rho_* =\tfrac{1}{2}(\ketbra{\epsilon_k}+\ketbra{\epsilon_{d+1-k}}).}
\end{equation}
Its initial and final energies under reversal both equal $\epsilon_*$.
For every $E\in[\epsilon_{\min},\epsilon_*]$, a mixture of $\rho_*$ and $\ketbra{\epsilon_1}$ has initial mean energy $E$ and final mean energy at least $E$.
Thus $E-R_{\mathrm{rev}}(E)\leq0$ there; at $E=\epsilon_*$ it is exactly zero.
The scalar-Hamiltonian case is immediate.
Finally, the maximum of the paired sums is at least their average, $2\epsilon_{\mathrm{mean}}$, so $\epsilon_*\geq\epsilon_{\mathrm{mean}}$.
\end{proof}

For completeness, the remaining classical optimization is
\begin{equation}
\label{sm:eq:Rrev-lp}
{R_{\mathrm{rev}}(E)=\max_{\substack{p_k\geq0,\ \sum_kp_k=1\\
                                  \sum_kp_k\epsilon_k=E}}
 \sum_kp_k\epsilon_{d+1-k}.}
\end{equation}
It is the upper boundary of the convex hull of $(\epsilon_k,\epsilon_{d+1-k})$; it need not be a single straight line above the threshold.
An extreme optimum uses at most two occupied levels.
For $\epsilon_i<\epsilon_j$ bracketing $E$, its candidate value is
\begin{equation}
{R_{ij}(E)=\frac{\epsilon_j-E}{\epsilon_j-\epsilon_i}\epsilon_{d+1-i}
 +\frac{E-\epsilon_i}{\epsilon_j-\epsilon_i}\epsilon_{d+1-j}.}
\end{equation}
Take the maximum of these candidates and the single-level values $\epsilon_{d+1-i}$ with $E=\epsilon_i$.
This also handles degeneracies without dividing by zero.
The two occupied levels and their interpolation weights give an explicit worst-case diagonal state for reversal.

\section{Achievability: special cases}
\label{sm:sec:achievability}

\subsection{Diagonal states and random unitaries (Proof of Theorem~4)}
\label{sm:sec:th4}
We use $\Delta$ for states diagonal in one fixed energy eigenbasis:
\begin{equation}
\label{sm:eq:fixed-diagonal-set}
 \Delta=\left\{\sum_kp_k\ketbra{\epsilon_k}:p_k\geq0,\ \sum_kp_k=1\right\}.
\end{equation}
For a nondegenerate Hamiltonian this is equivalent to $[\rho,\hat{H}]=0$.
For degenerate energies, commutation alone allows additional coherences inside energy eigenspaces and is a weaker condition.
The diagonal-input claim below refers to Eq.~\eqref{sm:eq:fixed-diagonal-set}.
Recall the two optimizations defined in the main text:
\begin{align}
\tilde{\mathcal{E}}_{\Delta}(E)
 &=\max_{\Lambda\in\mathcal{U}}\min_{\rho\in\sigma_E\cap\Delta}
 \Delta E(\rho,\Lambda),\\
 \tilde{\mathcal{E}}_{\mathcal{RU}}(E)
 &=\max_{\Lambda\in\mathcal{RU}}\min_{\rho\in\sigma_E}
 \Delta E(\rho,\Lambda).
\end{align}
\begin{theorem}
\label{sm:th:diag-and-rand}
For every admissible $E$, the minimum-ergotropy bound is attained by a unitary on diagonal inputs and by a random unitary channel on arbitrary inputs.
\begin{equation}
\tilde{\mathcal{E}}_{\Delta}(E)
        = \tilde{\mathcal{E}}_{\mathcal{RU}}(E)
        = \mathcal{E}_{\min}(E).
\end{equation}
\end{theorem}

\subsubsection{Energy coefficients of a unitary}
Since $\mathcal{U}$ is not a convex set, we cannot straightforwardly use the minimax theorem for the diagonal-input problem.
We avoid this issue by observing that the extracted energy depends only on a vector of diagonal matrix elements, rather than the whole population map.

\begin{lemma}[Attainable energy coefficients]
\label{sm:lem:diagonal-orbit}
Let $\boldsymbol{\epsilon}=(\epsilon_1,\ldots,\epsilon_d)^{\mathsf{T}}$ and $a_k=\braket{\epsilon_k|\hat{U}^\dagger\hat{H}\hat{U}|\epsilon_k}$.
The set of attainable vectors $\boldsymbol{a}$ is
\begin{equation}
\label{sm:eq:energy-polytope}
{\mathcal{C}_{\epsilon}
 =\operatorname{conv}\{P\boldsymbol{\epsilon}:P\text{ a permutation matrix}\}.}
\end{equation}
\end{lemma}
\begin{proof}
For every unitary, $a_k=\sum_j\epsilon_j|U_{jk}|^2$.
The squared moduli form a doubly stochastic matrix, so the permutation decomposition~\cite{DUFOSSE2016108} implies $\boldsymbol{a}\in\mathcal{C}_{\epsilon}$.
For completeness, we give the converse diagonal-orbit argument (the Schur--Horn property).
Sort a target $\boldsymbol{x}\in\mathcal{C}_{\epsilon}$ and the spectrum in decreasing order, writing the latter as $\eta_1\geq\cdots\geq\eta_d$.
Then
\begin{equation}
{\sum_{i=1}^m x_i\leq\sum_{i=1}^m\eta_i\quad(m<d),\qquad
 \sum_{i=1}^d x_i=\sum_{i=1}^d\eta_i.}
\end{equation}
Choose $k$ with $\eta_k\geq x_1\geq\eta_{k+1}$.
A rotation between the corresponding eigenvectors produces a unit vector $v_1$ with expectation $x_1$.
The compression of $\hat{H}$ to $v_1^\perp$ has the decreasingly ordered spectrum
\begin{equation}
{\eta_1,\ldots,\eta_{k-1},\quad
 \eta_k+\eta_{k+1}-x_1,\quad\eta_{k+2},\ldots,\eta_d.}
\end{equation}
The remaining target $(x_2,\ldots,x_d)$ satisfies the corresponding partial-sum inequalities: for $m<k$, its first $m$ entries sum to at most $mx_1\leq\sum_{i=1}^m\eta_i$; for $m\geq k$, subtract $x_1$ from the original $(m+1)$-entry inequality.
Induction supplies an orthonormal basis of $v_1^\perp$ with these expectations.
Together with $v_1$, this gives the required diagonal.
If $x_1=\eta_1$, remove that eigenvector first; the scalar case is immediate.
Finally, undo the sorting permutation of $\boldsymbol{x}$.
Thus every vector in $\mathcal{C}_{\epsilon}$ is attained by a unitary.
This is an equivalence of energy coefficients, not of entire bistochastic and unitary-induced population maps.
\end{proof}

\subsubsection{Diagonal inputs}
\begin{proof}[Proof of the first equality in Theorem~\ref{sm:th:diag-and-rand}]
Represent an input by its populations $\boldsymbol{p}\in\mathcal{P}_E\coloneqq \{\boldsymbol{p}\geq0:\sum_kp_k=1,\ \boldsymbol{p}\cdot\boldsymbol{\epsilon}=E\}$.
For a diagonal state, the extracted energy is $E-\boldsymbol{p}\cdot\boldsymbol{a}$.
Lemma~\ref{sm:lem:diagonal-orbit} therefore allows us to replace the optimization over unitaries by one over the compact convex set $\mathcal{C}_{\epsilon}$.
The payoff is continuous and affine in each argument, and $\mathcal{P}_E$ is also compact and convex.
The minimax theorem~\cite{sion1958minimax} then gives
\begin{equation}
{\begin{aligned}
\tilde{\mathcal{E}}_{\Delta}(E)
 &=\max_{\boldsymbol{a}\in\mathcal{C}_{\epsilon}}
   \min_{\boldsymbol{p}\in\mathcal{P}_E}(E-\boldsymbol{p}\cdot\boldsymbol{a})\\
 &=\min_{\boldsymbol{p}\in\mathcal{P}_E}
   \max_{\boldsymbol{a}\in\mathcal{C}_{\epsilon}}(E-\boldsymbol{p}\cdot\boldsymbol{a})
 =\min_{\rho\in\sigma_E\cap\Delta}\mathcal{E}(\rho)
 =\mathcal{E}_{\min}(E).
\end{aligned}}
\end{equation}
The last equality follows from Sec.~\ref{sm:sec:prop1}.
\end{proof}

\subsubsection{Random unitary channels}
\begin{proof}[Proof of the second equality in Theorem~\ref{sm:th:diag-and-rand}]
A random unitary channel has the form $\Lambda(\rho)=\sum_jp_j\hat{U}_j\rho\hat{U}_j^\dagger$.
Contrary to the set of unitary channels, the set $\mathcal{RU}$ is convex.
It is also compact in finite dimension, being the convex hull of a compact set.
The same holds for $\sigma_E$, which is an affine slice of the compact convex set of density matrices.
The function $\Delta E(\rho,\Lambda)$ is continuous and affine in each argument.
It follows that all hypotheses of the minimax theorem~\cite{sion1958minimax} are fulfilled, and we can exchange the minimum and maximum:
\begin{equation}
\begin{aligned}
\tilde{\mathcal{E}}_{\mathcal{RU}}(E)
 &=\max_{\Lambda\in\mathcal{RU}}\min_{\rho\in\sigma_E}\Delta E(\rho,\Lambda)
 =\min_{\rho\in\sigma_E}\max_{\Lambda\in\mathcal{RU}}\Delta E(\rho,\Lambda)\\
 &=\min_{\rho\in\sigma_E}\mathcal{E}(\rho)=\mathcal{E}_{\min}(E).
\end{aligned}
\end{equation}
For a fixed state, maximizing a linear payoff over mixtures of unitaries gives the same value as maximizing over the constituent unitaries, which justifies the penultimate equality.
\end{proof}
There is also a direct construction linking the two settings.
If $\hat{U}_E$ is optimal for diagonal inputs, apply the channel $\Lambda_E(\rho)=\hat{U}_E\Delta(\rho)\hat{U}_E^\dagger$.
Complete dephasing is the finite random unitary channel
\begin{equation}
\Delta(\rho)=\frac{1}{d}\sum_{j=0}^{d-1}D_j\rho D_j^\dagger,
 \qquad
 D_j=\sum_{k=1}^d e^{2\pi i j(k-1)/d}\ketbra{\epsilon_k}.
\end{equation}
It preserves the initial energy and removes sensitivity to unknown coherences, so $\Lambda_E$ guarantees at least $\mathcal{E}_{\min}(E)$.
The guarantee refers to the mean energy change of the averaged channel, not to the result of every individual random choice.

\subsection{Antisymmetric Hamiltonians (Proof of Theorem~5)}
\label{sm:sec:th5}
\begin{theorem}
\label{sm:th:anti-h}
Let $\hat{H}_{\mathrm{as}} = \sum_{k=1}^d \epsilon_k \ketbra{\epsilon_k}$ be an antisymmetric Hamiltonian.
Then the max-WCEE is equal to the minimum ergotropy and can be written explicitly as
\begin{equation}
\label{sm:eq:min-erg-antih}
\tilde{\mathcal{E}}(E)
        = \mathcal{E}_{\min}(E)
        = \begin{cases}
                0 & \text{if }E \leq \epsilon_{\mathrm{mean}}, \\
                2(E-\epsilon_{\mathrm{mean}}) & \text{if } E \geq \epsilon_{\mathrm{mean}}.
            \end{cases}
\end{equation}
\end{theorem}
\begin{proof}
For an antisymmetric spectrum, $\epsilon_k+\epsilon_{d+1-k}=c$ for all $k$.
Averaging over $k$ gives $c=2\epsilon_{\mathrm{mean}}$, and hence
\begin{equation}
\hat{U}_{\mathrm{rev}}^\dagger\hat{H}_{\mathrm{as}}\hat{U}_{\mathrm{rev}}
 =2\epsilon_{\mathrm{mean}}\mathbb{I}_d-\hat{H}_{\mathrm{as}}.
\end{equation}
Therefore, population reversal extracts exactly $2(E-\epsilon_{\mathrm{mean}})$ from any state in $\sigma_E$.
For $E\geq\epsilon_{\mathrm{mean}}$ this is nonnegative and Theorem~\ref{sm:th:general-unitary} identifies it with the max-WCEE.
For lower energies the identity is optimal and guarantees zero.

It remains to show that the ergotropic bound is saturated.
All antipassive states are optimally discharged by population reversal.
Thus, when $E\geq\epsilon_{\mathrm{mean}}$, every antipassive state in $\sigma_E$ has ergotropy $2(E-\epsilon_{\mathrm{mean}})$.
Such states exist throughout this interval, for example as mixtures of $\mathbb{I}_d/d$ and a highest-energy eigenstate.
Proposition~\ref{sm:prop:min-states} then gives $\mathcal{E}_{\min}(E)=2(E-\epsilon_{\mathrm{mean}})$.
For $E\leq\epsilon_{\mathrm{mean}}$, passive states exist and the minimum is zero.
This proves the stated equality on the full admissible interval.
\end{proof}

\section{The qutrit example}
\label{sm:sec:qutrit}
We now work out the calculations for the qutrit of the main text,
\begin{equation}
{\epsilon_1=0,\qquad \epsilon_2=\epsilon(1+\delta),\qquad
 \epsilon_3=2\epsilon,\qquad \epsilon>0,\quad -1<\delta<0.}
\end{equation}
Theorem~\ref{sm:th:general-unitary} determines the unitary optimum exactly, allowing a direct comparison of the protocols below.

\subsection{Energy-constrained minimum ergotropy}
The vertices of the antipassive simplex are
\begin{equation}
{\rho_1^\uparrow=\frac{\mathbb{I}_3}{3},\qquad
 \rho_2^\uparrow=\frac{1}{2}(\ketbra{\epsilon_2}+\ketbra{\epsilon_3}),\qquad
 \rho_3^\uparrow=\ketbra{\epsilon_3}.}
\end{equation}
Their energy--ergotropy coordinates are
\begin{equation}
{(\epsilon_{\mathrm{mean}},0),\qquad
 \left(\frac{\epsilon(3+\delta)}{2},\epsilon\right),\qquad
 (2\epsilon,2\epsilon),\qquad
 \epsilon_{\mathrm{mean}}=\epsilon\left(1+\frac{\delta}{3}\right).}
\end{equation}
For $\delta<0$, the middle point lies above the line connecting the other two.
Thus, throughout $E\in[\epsilon_{\mathrm{mean}},2\epsilon]$, a minimum-ergotropy state is a convex combination of the completely mixed state and the maximally excited state:
\begin{gather}
{\rho_E=(1-p)\frac{\mathbb{I}_3}{3}+p\ketbra{\epsilon_3},\qquad
 p=\frac{E-\epsilon_{\mathrm{mean}}}{2\epsilon-\epsilon_{\mathrm{mean}}},} \\
\label{sm:eq:qutrit-minimum}
{\mathcal{E}_{\min}(E)=
 \begin{cases}
 0,&0\leq E\leq\epsilon_{\mathrm{mean}},\\
 \displaystyle\frac{6}{3-\delta}(E-\epsilon_{\mathrm{mean}}),
 &\epsilon_{\mathrm{mean}}\leq E\leq2\epsilon.
 \end{cases}}
\end{gather}
This is the red curve in Fig.~3~(b) of the main text.

\subsection{Optimal unitary extraction}
Population reversal simply swaps the ground state and maximally excited state:
\begin{equation}
{\hat{U}_{\mathrm{rev}}=\ketbra{\epsilon_3}{\epsilon_1}
 +\ketbra{\epsilon_2}+\ketbra{\epsilon_1}{\epsilon_3},\qquad
 \hat{H}_{\mathrm{rev}}=\operatorname{diag}(2\epsilon,\epsilon_2,0).}
\end{equation}
It does not mix populations and coherences in the energy variation.
The three points defining $R_{\mathrm{rev}}$ in Eq.~\eqref{sm:eq:Rrev-lp} are $(0,2\epsilon)$, $(\epsilon_2,\epsilon_2)$, and $(2\epsilon,0)$.
Since $\delta<0$, the middle point is below the line joining the other two.
Consequently,
\begin{equation}
{R_{\mathrm{rev}}(E)=2\epsilon-E,\qquad
 \min_{\rho\in\sigma_E}\Delta E(\rho,\Lambda_{\mathrm{rev}})=2(E-\epsilon),}
\end{equation}
and Theorem~\ref{sm:th:general-unitary} gives the exact optimum
\begin{equation}
\label{sm:eq:qutrit-unitary-optimum}
{\tilde{\mathcal{E}}(E)=\max\{0,2(E-\epsilon)\},\qquad
 \epsilon_*=\epsilon>\epsilon_{\mathrm{mean}}.}
\end{equation}
This is the blue curve wherever positive in Fig.~3~(b) of the main text, with zero extraction obtained by the identity below $\epsilon$.
The interval $(\epsilon_{\mathrm{mean}},\epsilon]$ exhibits positive state-aware extractable energy for every input, but no positive single-unitary guarantee.
Above $\epsilon$, the gap is
\begin{equation}
{\mathcal{E}_{\min}(E)-\tilde{\mathcal{E}}(E)
 =\frac{-2\delta}{3-\delta}(2\epsilon-E),\qquad
 \epsilon\leq E\leq2\epsilon.}
\end{equation}
It closes at the highest energy, where the input state is known.

\subsection{The optimal protocol for diagonal inputs}
We next characterize the unitary that guarantees extraction of the minimum ergotropy from every diagonal state of fixed mean energy $E\geq\epsilon_{\mathrm{mean}}$.
The minimum-ergotropy state $\rho_E$ is degenerate in its first two populations.
The optimal unitary can therefore be sought as a rotation of these two levels followed by population reversal:
\begin{equation}
\label{sm:eq:qutrit-phase-family}
{\hat{U}_{\Delta}(\phi)=
 \begin{pmatrix}
 0&0&1\\
 e^{i\phi}\sqrt{q}&\sqrt{1-q}&0\\
 e^{i\phi}\sqrt{1-q}&-\sqrt{q}&0
 \end{pmatrix},\qquad
 q=\frac{-4\delta}{(1-\delta)(3-\delta)}\in(0,1),}
\end{equation}
where $\phi$ is arbitrary.
Let $\Lambda_{\phi}(\rho)=\hat{U}_{\Delta}(\phi)\rho\hat{U}_{\Delta}^\dagger(\phi)$ denote the corresponding unitary channel.
This family obeys $\hat{U}_{\Delta}(\phi)=\hat{U}_{\Delta}(0)\operatorname{diag}(e^{i\phi},1,1)$; $\phi$ is a relative input phase, not a global phase.
To derive $q$, set $s=\epsilon_2/\epsilon_3=(1+\delta)/2$ and parametrize an arbitrary diagonal state at that energy as
\begin{equation}
{\begin{aligned}
\rho_{\Delta}
 &=\lambda_1\ketbra{\epsilon_1}+\lambda_2\ketbra{\epsilon_2}
   +\lambda_3\ketbra{\epsilon_3},\\
 \lambda_1&=\frac{1-p}{3}-(1-s)z,\qquad
 \lambda_2=\frac{1-p}{3}+z,\qquad
 \lambda_3=\frac{1+2p}{3}-sz.
\end{aligned}}
\end{equation}
Here $p=(E-\epsilon_{\mathrm{mean}})/(2\epsilon-\epsilon_{\mathrm{mean}})$ and $z$ ranges over the values for which the populations are nonnegative.
Rotating the first two levels gives intermediate populations $\tilde{\lambda}_1=(1-q)\lambda_1+q\lambda_2$ and $\tilde{\lambda}_2=q\lambda_1+(1-q)\lambda_2$.
The extracted energy after reversal is therefore
\begin{equation}
{\begin{aligned}
\Delta E(\rho_{\Delta},\Lambda_{\phi})
 &=\epsilon_3[\lambda_3-\lambda_1+q(1-s)(\lambda_1-\lambda_2)]\\
 &=\epsilon_3p+\epsilon_3z[(1-2s)-q(1-s)(2-s)].
\end{aligned}}
\end{equation}
Choosing $q=(1-2s)/[(1-s)(2-s)]$, which is precisely Eq.~\eqref{sm:eq:qutrit-phase-family}, removes the dependence on $z$.
We then extract $\epsilon_3p=\mathcal{E}_{\min}(E)$ from every compatible diagonal state.
This proves optimality through the converse bound.

\subsection{Unknown coherences and worst-case performance}
For arbitrary inputs, compute the energy observable directly:
\begin{equation}
{\hat{U}_{\Delta}^\dagger(\phi)\hat{H}\hat{U}_{\Delta}(\phi)=
 \begin{pmatrix}
 \displaystyle\frac{2\epsilon(3+\delta)}{3-\delta}
 &-\tfrac{1}{2}C_\delta e^{-i\phi}&0\\
 -\tfrac{1}{2}C_\delta e^{i\phi}
 &\displaystyle\frac{\epsilon(1-\delta)(3+\delta)}{3-\delta}&0\\
 0&0&0
 \end{pmatrix},\qquad
 C_\delta=2\epsilon(1-\delta)\sqrt{q(1-q)}.}
\end{equation}
It follows that, with $\rho_{12}=\braket{\epsilon_1|\rho|\epsilon_2}$,
\begin{equation}
\label{sm:eq:qutrit-coherence}
{\Delta E(\rho,\Lambda_{\phi})
 =\mathcal{E}_{\min}(E)+C_\delta\operatorname{Re}(e^{i\phi}\rho_{12}),
 \qquad E\geq\epsilon_{\mathrm{mean}}.}
\end{equation}
For any fixed $\phi$, the unknown phase of a nonzero $\rho_{12}$ can make the second term negative.
Positivity of $\rho$ implies $|\rho_{12}|\leq\sqrt{\lambda_1\lambda_2}$.
The energy constraint gives
\begin{equation}
{2\epsilon\lambda_1+\epsilon(1-\delta)\lambda_2=2\epsilon-E,
 \qquad
 \sqrt{\lambda_1\lambda_2}
 \leq\frac{2\epsilon-E}{2\epsilon\sqrt{2(1-\delta)}}.}
\end{equation}
For $E\in[\epsilon_{\mathrm{mean}},2\epsilon]$ the bound is attained by
\begin{equation}
{\lambda_1=\frac{2\epsilon-E}{4\epsilon},\qquad
 \lambda_2=\frac{2\epsilon-E}{2\epsilon(1-\delta)},\qquad
 \lambda_3=1-\lambda_1-\lambda_2,}
\end{equation}
which are nonnegative throughout this interval, and by the pure state
\begin{equation}
{\ket{\psi_{\phi}}=\sqrt{\lambda_1}\ket{\epsilon_1}
 -e^{i\phi}\sqrt{\lambda_2}\ket{\epsilon_2}
 +\sqrt{\lambda_3}\ket{\epsilon_3}.}
\end{equation}
Thus the exact worst-case performance of the diagonal-optimal protocol is
\begin{gather}
{W_{\Delta}(E)\coloneqq\min_{\rho\in\sigma_E}
 \Delta E(\rho,\Lambda_{\phi})
 =\frac{6(E-\epsilon_{\mathrm{mean}})-r_\delta(2\epsilon-E)}{3-\delta},} \\
{r_\delta\coloneqq
 \sqrt{\frac{-2\delta(3+\delta^2)}{1-\delta}},\qquad
 \tilde{\epsilon}=\frac{2\epsilon(3+\delta+r_\delta)}{6+r_\delta}.}
\end{gather}
The green curve in Fig.~3~(b) of the main text displays $W_{\Delta}(E)$ where positive.
It is independent of $\phi$ and becomes positive only above $\tilde{\epsilon}>\epsilon$, whereas the unrestricted optimum already becomes positive above $\epsilon$.
At $E=2\epsilon$, all three values coincide.

\subsection{Random unitary extraction}
The preceding expression makes the role of random unitaries explicit.
Each member is optimal for diagonal states, but its coherent contribution changes sign when $\phi$ is shifted by $\pi$.
Therefore the random unitary channel
\begin{equation}
{\Lambda_{\mathcal{RU}}=\tfrac{1}{2}(\Lambda_{\phi}+\Lambda_{\phi+\pi})}
\end{equation}
satisfies
\begin{equation}
{\begin{aligned}
\Delta E(\rho,\Lambda_{\mathcal{RU}})
 &=\tfrac{1}{2}[\mathcal{E}_{\min}(E)+C_\delta\operatorname{Re}(e^{i\phi}\rho_{12})]
  +\tfrac{1}{2}[\mathcal{E}_{\min}(E)-C_\delta\operatorname{Re}(e^{i\phi}\rho_{12})]\\
 &=\mathcal{E}_{\min}(E)
\end{aligned}}
\end{equation}
for every $\rho\in\sigma_E$ and every $\phi$, when $E\geq\epsilon_{\mathrm{mean}}$.
Two relative phases suffice; a continuous phase average and complete dephasing of the output are unnecessary.
For $E\leq\epsilon_{\mathrm{mean}}$ the identity achieves the optimal zero guarantee.
These guarantees concern the mean energy change averaged over random realizations.

\putunitbibliography

\end{bibunit}
\end{document}